\documentclass[12pt,preprint]{aastex}
\usepackage{amssymb, amsmath}

\shorttitle{Determining the Physical Parameters of a New Type of Wolf-Rayet Star}

\shortauthors{Neugent et al.}

\begin{document}

\title{The Evolution and Physical Parameters of WN3/O3s:\\ a New Type of Wolf-Rayet Star\altaffilmark{*}}

\author{Kathryn F.\ Neugent\altaffilmark{1,2}, Philip Massey\altaffilmark{1,2}, D. John Hillier\altaffilmark{3}, and Nidia Morrell\altaffilmark{4}}

\altaffiltext{*}{This paper includes data gathered with the 6.5-m Magellan Telescopes located at Las Campanas Observatory, Chile. It is additionally based on observations made with the NASA/ESA {\it Hubble Space Telescope}, obtained at the Space Telescope Science Institute, which is operated by the Association of Universities for Research in Astronomy, Inc., under NASA contract NAS 5-26555. These observations were associated with program GO-13780.}
\altaffiltext{1}{Lowell Observatory, 1400 W Mars Hill Road, Flagstaff, AZ 86001; kneugent@lowell.edu; phil.massey@lowell.edu}
\altaffiltext{2}{Department of Physics and Astronomy, Northern Arizona University, Flagstaff, AZ, 86011-6010}
\altaffiltext{3}{Department of Physics and Astronomy, University of Pittsburgh, Pittsburgh, PA 15260; hillier@pitt.edu}
\altaffiltext{4}{Las Campanas Observatory, Carnegie Observatories, Casilla 601, La Serena, Chile; nmorrell@lco.cl}

\begin{abstract}
As part of a search for Wolf-Rayet (WR) stars in the Magellanic Clouds, we have discovered a new type of WR star in the Large Magellanic Cloud (LMC). These stars have both strong emission lines, as well as He\,{\sc ii} and Balmer absorption lines and spectroscopically resemble a WN3 and O3V binary pair. However, they are visually too faint to be WN3+O3V binary systems. We have found nine of these WN3/O3s, making up $\sim6$\% of the population of LMC WRs. Using {\sc cmfgen}, we have successfully modeled their spectra as single stars and have compared the physical parameters with those of more typical LMC WNs. Their temperatures are around 100,000 K, a bit hotter than the majority of WN stars (by around 10,000 K) although a few hotter WNs are known. The abundances are what you would expect for CNO equilibrium. However, most anomalous are their mass-loss rates which are more like that of an O-type star than a WN star. While their evolutionary status is uncertain, their low mass-loss rates and wind velocities suggest that they are not products of homogeneous evolution. It is possible instead that these stars represent an intermediate stage between O stars and WNs. Since WN3/O3 stars are unknown in the Milky Way, we suspect that their formation depends upon metallicity, and we are investigating this further by a deep survey in M33, which posses a metallicity gradient.
\end{abstract}


\section{Discovery}
Spurred on by the discovery of a previously unknown WO-type Wolf-Rayet (WR) star in the Large Magellanic Cloud (LMC) by Neugent et al. (2012a), we undertook a multi-year search for WRs in the Magellanic Clouds. So far we have imaged the entire optical disks of the Magellanic Clouds and discovered 16 new WRs (Massey et al.\ 2014, 2015, 2017). However, it is not the {\it number} of newly discovered WRs that makes our survey so successful, but the {\it types} of these stars. Nine of them, or 6\% of the known LMC WRs, and nearly 10\% of the known LMC WNs appear to belong to an entirely new class of WRs. Given their statistical significance, they may represent some previously unrecognized part of a WR's evolution at intermediate (and lower?) metallicities. We have additionally discovered a tenth star that is similar, but not identical, to the other nine stars. It will be discussed in a separate paper.

\begin{figure}
\epsscale{.5}
\plotone{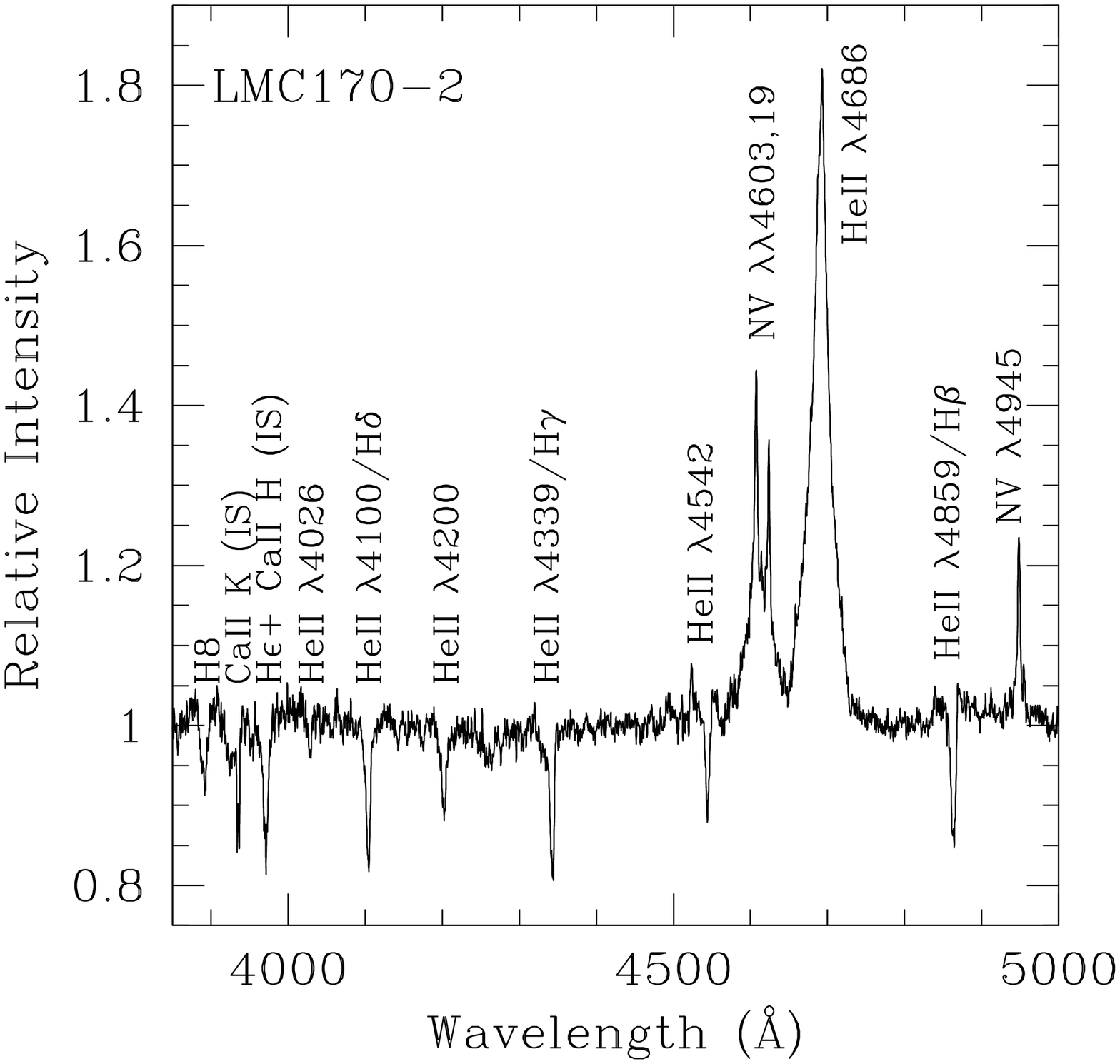}
\caption{\label{fig:spectra} Spectrum of LMC170-2, one of our newly discovered WN3/O3s. The WN3 classification comes from the star's N\,{\sc v} emission ($\lambda\lambda$4603,19 and $\lambda$4945), but lack of N\,{\sc iv}. The O3V classification comes from the strong He\,{\sc ii} absorption lines but lack of He\,{\sc i}.}
\end{figure}

Figure~\ref{fig:spectra} shows the spectrum of one of our newly discovered stars, LMC170-2, first discussed in Massey et al.\ (2014). Based on this optical spectrum, one might initially classify this star as a WN3+O3V binary. The star's N\,{\sc v} emission ($\lambda\lambda$4603,19 and $\lambda$4945), but lack of N\,{\sc iv}, suggests a WN3-type, while strong He\,{\sc ii} absorption lines but lack of He\,{\sc i} are characteristic of an O3V. However such a WN3+O3V pairing is improbable. First, O3Vs are the hottest and most luminous of the dwarfs, making them quite rare. In the LMC, only around a dozen are known outside of the 30 Dor region (Skiff 2014), so the probability of discovering nine more of them paired with WN3s is extremely low. Second, the lifetimes of these stars preclude such a pairing if one assumes single star evolution (a lack of radial velocity variations suggests these stars are single) and that no interactions have taken place: WN3s take $\sim$3-5 million years to form whereas an O3V only lasts a million years. But unlikely pairings aside, an O star cannot be present. For one, the most prominent UV line in O stars, C\,{\sc iv} $\lambda$1550, is absent in our UV data. But most convincingly, the WN3/O3Vs are faint with M$_V \sim -2.5$. O3V stars, on the other hand, have much brighter absolute magnitudes with M$_V \sim -5.5$ (Conti 1988). For these reasons we conclude our newly discovered WRs are not WN3+O3V binaries and adopt the naming convention of WN3/O3 to describe the composite appearance of the spectra.

As discussed by Massey et al.\ (2014) and Neugent et al. (2015), we successfully modeled a moderate resolution optical spectrum of LMC170-2 using a single set of physical parameters. Here we examine how well this model fits our newly obtained {\it HST}/COS UV and Magellan/FIRE NIR data for this star. We then extend our modeling efforts to the remaining eight stars, two of which also have COS/UV data. Finally, we discuss what these results suggest about the origin and nature of these newly found WRs. 

\section{Observations and Reductions}
We obtained medium-dispersion optical spectra for all nine WN3/O3s, high-dispersion optical spectra for one WN3/O3, UV spectra for three WN3/O3s, and NIR spectra for one WN3/O3. For LMC170-2, we obtained all four giving us coverage from $1000-25000$ \AA. A summary of the observations can be found in Table~\ref{tab:observations} and broadband photometry values can be found in Table~\ref{tab:photometry}.

\subsection{Optical}
The nine WN3/O3s were first spectroscopically confirmed using Magellan Echellette (MagE, see Marshall et al.\ 2008) on the 6.5-m Magellan telescopes (MagE moved from the Clay to the Baade in October 2015). MagE has a wavelength coverage of $3180-9400$ \AA\ and a resolving power of 4100. Observations were taken between Oct 2013 and Jan 2016 with clear skies and the seeing, as reported on the guide camera, was between 0\farcs6 and 1\arcsec. Further details about our observing procedures and data reduction techniques can be found in Massey et al.\ (2014, 2015).

The resolution of the MagE spectra (75 km s$^{-1}$) is good, but we wanted a higher resolution spectrum to confirm the rotational velocity for LMC170-2 (120-150 km~s$^{-1}$, as discussed later). We therefore obtained a high dispersion optical spectrum ($R \sim 12000$ in the red and $\sim 17000$ in the blue translating to $\sim 24$ and $\sim 18$ km s$^{-1}$, respectively) using Magellan Inamori Kyocera Echelle (MIKE), a high-throughput double echelle spectrograph on the 6.5-m Magellan Clay telescope. The spectrum was taken on (UT) 2 Sep 2014 through some cirrus with $\sim$1\arcsec\ seeing using the 2\arcsec\ slit. A binning of $2\times2$ was used. We used a wide slit because the position angle of the slit is fixed, there is no atmospheric dispersion corrector, and scheduling constraints precluded our observing the star near the parallactic angle. The data was reduced using a combination of the $mtools$ package and the standard IRAF\footnote{IRAF is distributed by the National Optical Astronomy Observatory, which is operated by the Association of Universities for Research in Astronomy (AURA) under cooperative agreement with the National Science Foundation.} echelle routines, similar to what is described by Massey et al.\ (2012).

The fluxes of the optical spectrophotometry were adjusted to the broad-band V magnitudes of Zaritsky et al. (2004) in order to correct for differential slit losses between the spectrophotometric standards and the observations due to seeing variations. 

\subsection{Ultraviolet}
We were awarded 6 orbits in Cycle 22 (program GO-13780) to observe LMC079-1, LMC170-2, and LMC277-2, in the far-UV (FUV) using COS. Our objects were too bright for acquisition through the Primary Science Aperture (PSA) with Mirror A, so we performed the acquisition using dispersed light and the G140L grating with the 1105 \AA\ setting. The observations were obtained with the G140L grating with the two standard wavelength settings at 1105 \AA\ and 1280 \AA\ for complete wavelength coverage in the FUV ($\sim$900 \AA\ to 2000 \AA) with a resolving power $R\sim 2000$. The G140L/1280 setting was observed with Segment A and B of the detector, while the G140L/1105 setting was observed only with Segment A, as this setting requires the voltage on Segment B to be lowered to avoid dangerously high count rates from zeroth-order light (see Debes et al. 2015). 

Fixed pattern noise limits the signal-to-noise that can be achieved with COS to 15-25 per resolution element in the FUV; to do better than this, we followed the recommendations of the COS team and observed with the grating tweaked by four different grating steps for each nominal wavelength setting. These shifts were automatically corrected as part of the standard pipeline reductions. We did find that a significant wavelength shift (3 pixels) was needed to combine the overlapping data from the two wavelength settings; correspondence with the help desk revealed the surprising information that this pixel shift was within expectations, which could be as large as 10~pixels. 

\subsection{Near Infrared}
In addition to the optical and UV observations, we observed LMC170-2 using the Folded port InfraRed Echellette (FIRE) spectrograph in high dispersion mode on the 6.5-m Magellan Baade telescope on (UT) 3 Feb 2015. FIRE covers the full 0.8-2.5$\micron$ band at a resolving power of $R\sim 6000$, giving us coverage of LMC170-2 from $1000-25000$ \AA\ (including the COS and MagE spectra). However, LMC170-2 was observed under less than optimal conditions (airmass of $\sim$2), with modest seeing (0\farcs8 - 0\farcs9 in the optical as measured on the guide camera) using the 0\farcs6 slit. The observation consisted of four 602.4 second observations at slit positions ``A" and ``B" in an ABBA pattern using the ``sample-up-the-ramp" read-mode. The high-gain (1.3 e/DN) setting was used. A ThAr calibration exposure was made immediately afterwards.  A telluric standard, HD 47925, was observed at a similar airmass (2.08-2.18) in four 30 second exposures in an ABBA pattern immediately afterwards using the ``Fowler1" read-mode.  The data were reduced using the standard FIRE pipeline reduction kindly provided by Robert Simcoe and the FIRE team.

\section{{\sc cmfgen} Modeling Overview}
As is indicative of stars approaching the Eddington Limit, a WR's spectrum is heavily influenced by strong winds and high mass-loss rates. Keeping the luminosity near, but below, the Eddington limit can make modeling WRs quite a challenge. Additionally, the stars' high surface temperatures mean that the assumption of local thermal equilibrium (LTE) is no longer valid. Instead, the high degree of ionization (and decreased opacity) causes the radiation field to decouple from the local electron temperature. Furthermore, WR atmospheres are significantly extended when compared to their radius. Thus, plane-parallel geometry cannot be used and instead spherical geometry must be included. Finally WR models must be fully blanketed and include the effects of thousands of overlapping metal lines, which occur at the (unobservable) short wavelengths ($< 1000$ \AA) where most of the flux of the star is produced. Two codes are currently capable of including these complexities, the Potsdam Wolf-Rayet Models, or PoWR (Gr\"{a}fener et al.\ 2002), and the CoMoving Frame GENeral spectrum analysis code, {\sc cmfgen} (Hillier \& Miller 1998). We employ the latter here.

To model a star with {\sc cmfgen} there are a number of parameters that the user can vary (somewhat) independently. When fitting the WN3/O3s, we initially fit the $\beta$ parameter, which characterizes the stellar wind law (Castor \& Lamers 1979, Santolaya-Rey et al., 1997), and a filling factor for the stellar wind clumping law. Next we varied the following in an attempt to find the best model fit to our spectra: the effective temperature $T_{\rm eff}$, the logarithm of the effective surface gravity $\log g_{\rm eff}$, the mass-loss rate $\dot{M}$, the terminal velocity $V_\infty$, the ratio of He to H, and the chemical abundances (specifically carbon, nitrogen and oxygen; or CNO, but it is possible to set the values of Ne, Mg, Si, P, S, Cl, Ar, and Fe as well). Finally, we used the absolute visual magnitude to fix the radius using the reddened model spectral energy distribution (SED), as described in Section 5.2. We then re-ran the model, resulting in a new luminosity. We continued to adjust the parameters and re-run the models until the model matched the observed spectrum.

While modeling these stars is complex, there are a few basic guidelines we followed. Since their spectra is dominated by stellar winds, we first varied $\beta$ to fit the shape of H$\alpha$. Since varying $\beta$ drastically varies the derived values for other stellar wind parameters, we settled on $\beta = 1$ for all of our stars\footnote{Our initial model of LMC170-2 reported by Massey et al.\ (2014) used $\beta=0.8$ but we later settled on a value of $\beta=1$ to best match the spectra.}. This allowed us to perform a more direct comparison of the remaining stellar wind parameters. After setting $\beta$ to 1, we then fit the terminal velocity initially using H$\alpha$ and then using the blue-ward portion of the P Cygni profile of both O\,{\sc vi} $\lambda$1038 and N\,{\sc v} $\lambda$1240 (for stars with UV data). Finally, we used He\,{\sc ii} $\lambda$4686 and H$\alpha$ to approximate the mass-loss rate. Once we had a good sense of the stellar wind-dominated parameters we went on to fit the He\,{\sc i} and He\,{\sc ii} lines (or lack thereof) by varying the effective temperature, the wings of the Balmer lines by varying the surface gravity, and the strengths of the He\,{\sc i} and He\,{\sc ii} lines by varying the He/H number ratio. Finally, we varied the abundances (primarily nitrogen) based on the strongest CNO lines. In practice, modeling is an iterative process since varying one parameter will often require a change in another parameter in order to maintain the fit quality.

Though we can discuss fitting the lines of these stars as an iterative process, it is nearly impossible to fit all of the lines in this way without making compromises. This is due to a few reasons. First is the coupling between different parameters. For normal WNE stars, determining the key parameters (effective temperature, mass-loss rate, etc.) for an assumed velocity law is very straightforward. However, in the current analysis, these physical parameters cannot be determined independently in this way. Moreover, the presence of both photospheric and wind lines means that the resulting fits also have some sensitivity to the surface gravity, the adopted velocity law, and the filling factor (as well as its implementation). Secondly, in these models we are making the usual assumptions of a spherical and static flow, and we treat clumping using volume filling factors. The applicability of these assumptions, and their influence on the quality of the fits, is unknown.

\section{Modeling Efforts: LMC170-2}
The preliminary modeling (Massey et al. 2014; Neugent et al. 2015) was from a single MagE exposure of LMC170-2; we now have considerably higher signal-to-noise data, as well as the high dispersion MIKE optical data.  Finally, we have both UV and NIR for LMC170-2, making this the logical choice to refine our modeling work. Additionally, as is discussed below, all nine stars have very similar spectra (at least in the optical), so we were quite certain that once we obtained a satisfactory fit to LMC170-2, we could use that as a starting point to fit the remaining eight stars.

\subsection{Optical Data}
We derived our first set of properties using only the MagE optical data (the UV and NIR spectra had not yet been obtained), achieving a good visual fit to both the WN3-like emission components and the O3V-like absorption components. As Figure~\ref{fig:spectra} shows, there is strong He\,{\sc ii}, no He\,{\sc i}, N\,{\sc v}, and no N\,{\sc iv}. While the initial fit was presented in Massey et al.\ (2014), here we describe in detail how we determined the final parameters for LMC170-2.

We first modeled the stellar wind assuming a $\beta$ parameter of 0.8 and a stellar wind clumping law with a volume filling factor of 0.1 at $V_\infty$, as argued by Owocki \& Cohen (2008). The terminal velocity is generally measured using the P Cygni profile of the C\,{\sc iv} $\lambda$1550 doublet as the shortward line profile is known to be sensitive to the star's terminal velocity. However, without UV data, a value of 2400 km s$^{-1}$ fit the width of the optical emission lines well. Given this terminal velocity, we then fit He\,{\sc ii} $\lambda$4686 and H$\alpha$ and determined a mass-loss rate of $1.15 \times 10^{-6}$ $\dot{M}$ yr$^{-1}$ ($\log \dot{M} = -5.94$ dex). We used a turbulent velocity of 15 km s$^{-1}$.

Next we constrained the effective temperature and surface gravity. As discussed above, this is generally done by fitting the He\,{\sc i} and He\,{\sc ii} lines. However, the lack of He\,{\sc i} in the spectrum of LMC170-2 dictates a high temperature, but also prevents us from determining the temperature using this method. The N\,{\sc iv} lines were absent for the same reason, also preventing us from using the strengths of the N\,{\sc iv} and N\,{\sc v} lines as temperature indicators. Based on the work of Hainich et al.\ (2014), who modeled WN stars in the LMC using PoWR, a minimum temperature for a LMC WN3 is $\sim$70K. We used this as a starting point, knowing that the temperature was likely much higher. To fit both the He\,{\sc ii}, N\,{\sc v}, and the non-existent He\,{\sc i} and N\,{\sc iv}, we increased the temperature to 100K. A high surface gravity was needed to keep the star in hydrostatic equilibrium following the linear relationship between the effective temperature and $\frac{\dot{M}}{\sqrt{fR^3}}$ (where $f$ is the filling factor, taken to be 0.1 and $R$ is the radius, discussed later) (Schmutz et. al 1989). Using this relationship, we derived a surface gravity of between 4.85 - 4.95 dex, which additionally fit the wings of the Balmer lines well.

We first adopted one-half solar values for the LMC chemical abundances, corresponding to what we expect for the initial metallicity of LMC massive stars. By the time a massive star reaches the WN phase, the star will have enhanced N at the expense of C and O. A N abundance of 10$\times$ solar (0.011 by mass) was derived primarily from the strengths of N\,{v} $\lambda \lambda$4603,19 and N\,{v} $\lambda$4944. C and O abundances 0.05$\times$ solar ($1.53 \times 10^{-4}$ and $4.77 \times 10^{-4}$ by mass, respectively) were derived from the lack of C\,{\sc iv} $\lambda \lambda$1548,1551 and C\,{\sc iv} $\lambda \lambda$5801,5812 as well as the strengths of O\,{\sc iv/v} $\lambda \lambda$6690-3. 

An un-evolved massive star will have a surface He/H number ratio of $\sim 0.1$. As the star evolves, the surface He/H ratio will increase as more hydrogen is converted into helium in the core and mixed to the surface. By the time a massive star first reaches the WN phase, it generally is assumed to have a He/H ratio of $> \sim 0.8$ (see, e.g. Georgy et al. 2012). Most early-type WNs are hydrogen poor with surface He/H ratios greater than 12 (Hainich et al. 2014)\footnote{Specifically, 40 (77\%) of the WN2-4 in Table 2 of Hainich et al.\ 2014 have no detectable hydrogen, while only 12 (13\%) do.}. By contrast, all of the WN3/O3s have an intermediate He/H ratio of $\sim 1$ suggesting that they are still unusually hydrogen rich for an early-type WN. The implications of this are discussed later.

To measure the projected rotational velocity, we relied on how well the spectra fit the N\,{\sc v} lines. We were primarily concerned with fitting the shape of the line, as opposed to fitting the strength, because the strength could easily be changed by altering either the N abundance or the effective temperature. A projected rotational velocity of 150$\pm$10 km s$^{-1}$ fit our spectra well. We expect that there is a minor instrumental broadening component to the 150 km s$^{-1}$ $v \sin{i}$ value that we have been using in modeling the spectrum. However, we have compared our MagE spectrum of LMC170-2 with our higher resolution ($R=12000-17000$) MIKE spectrum, and see no appreciable difference in the line widths.

We additionally performed a sensitivity analysis by varying each parameter until there was a clear discrepancy between the observed spectrum and the model. The uncertainty in our modeling is then the value of this change. The values we found are consistent with the uncertainties found by Martins et al. (2013) when modeling LMC WN stars using {\sc cmfgen}. It should be noted that these uncertainties represent the uncertainty within the particular model and not within the modeling code itself. We later discuss the uniqueness of these stars as a whole, or the parameter range that describes the WN3/O3s.

\begin{figure}
\epsscale{0.4}
\plotone{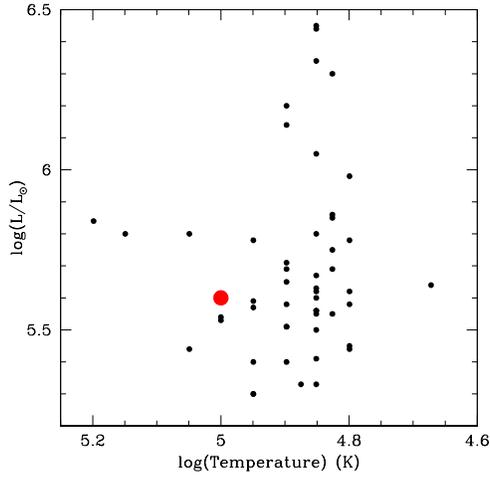}
\caption{\label{fig:tefflum} The bolometric luminosity vs.\ temperature for LMC170-2. The small black circles represent the early-type LMC WNs (WN3s and WN4s) analyzed by Hainich et al.\ (2014). LMC170-2 is represented as a large red circle. The bolometric luminosity of LMC170-2 is consistent with that of other early type LMC WNs but the temperature is a bit on the high side. In this plot, the temperature values from Hainich et al.\ (2014) refer to T* (the temperature at $\tau =20$ while the temperature of the WN3/O3 refers to T$_{\rm{eff}}$ (per definition at $\tau = 2/3$). While these two temperatures are generally quite close to one another, it should be noted that they do not have the same definition.}
\end{figure}

\begin{figure}
\epsscale{0.4}
\plotone{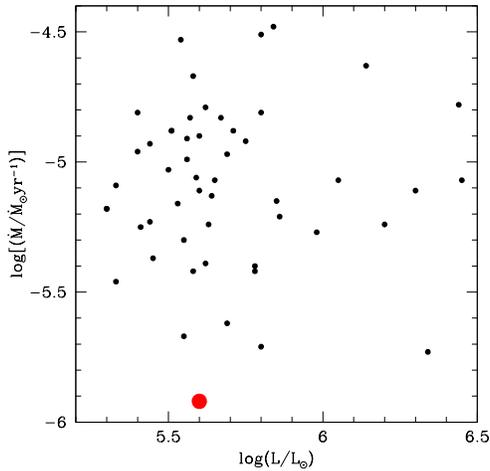}
\caption{\label{fig:massloss} The luminosity vs.\ mass-loss rate for LMC170-2. The small black circles represent the early-type LMC WNs (WN3s and WN4s) analyzed by Hainich et al.\ (2014). LMC170-2 is represented as a large red circle. The mass-loss rate of LMC170-2 is much lower than that of other early-type LMC WNs.}
\end{figure}

The physical properties of the model fit to LMC170-2 as well as the uncertainties are summarized in Table~\ref{tab:params}. The majority of these properties are comparable to those found for early-type LMC WNs (see Hainich et al. 2014). Despite the star's faint visual magnitudes, its bolometric luminosity is normal due to the very high effective temperature. While its temperature is a bit high for WNs, it is still within the range found for other WN3s however, the two known LMC WN2s have significantly higher temperatures, as is shown in Figure~\ref{fig:tefflum}. This star appears to be evolved, with significantly enriched N and He. While Hainich et al.\ (2014) used a N mass fraction of 0.004 for their modeling, we found that a value of 0.011 was needed to match the N lines (closer to the 0.008 N mass fraction used by Hamann et al. 2000). The most unusual value is its mass-loss rate. As Figure~\ref{fig:massloss} shows, our WN3/O3 star LMC170-2 has a mass-loss rate that is three to five times lower than that found for other early-type LMC WNs by Hainich et al.\ (2014). This low mass-loss rate appears to hold true for all nine WN3/O3s currently known. However, this matches the fact that the absorption lines aren't masked by emission due to strong stellar winds. We will discuss the implications of this in Section 6.

\subsection{UV and NIR}
Since the physical properties for this model were determined using only the optical data, we next investigated how well the model matched our UV and NIR spectra. 

Looking at the UV data, we first noticed the lack of C\,{\sc iv} $\lambda$1550, an O-type star's strongest line in the UV, further proving the lack of an O star. Without the C\,{\sc iv} $\lambda$1550 doublet, we relied on a combination of the other UV P Cygni lines to determine the terminal velocity. Based on our fits to the shortward line profile of the O\,{\sc vi} $\lambda$1038 resonance doublet and N\,{\sc v} $\lambda$1240, we found that increasing the terminal velocity from 2400 km s$^{-1}$ (based on fitting H$\alpha$) to 2800 km s$^{-1}$ yielded slightly better agreement to the UV lines. However, this increased terminal velocity then did not fit H$\alpha$. In the end, we settled on a compromise of 2600$\pm$200 km s$^{-1}$ for the terminal velocity.

\begin{figure}
\epsscale{0.32}
\plotone{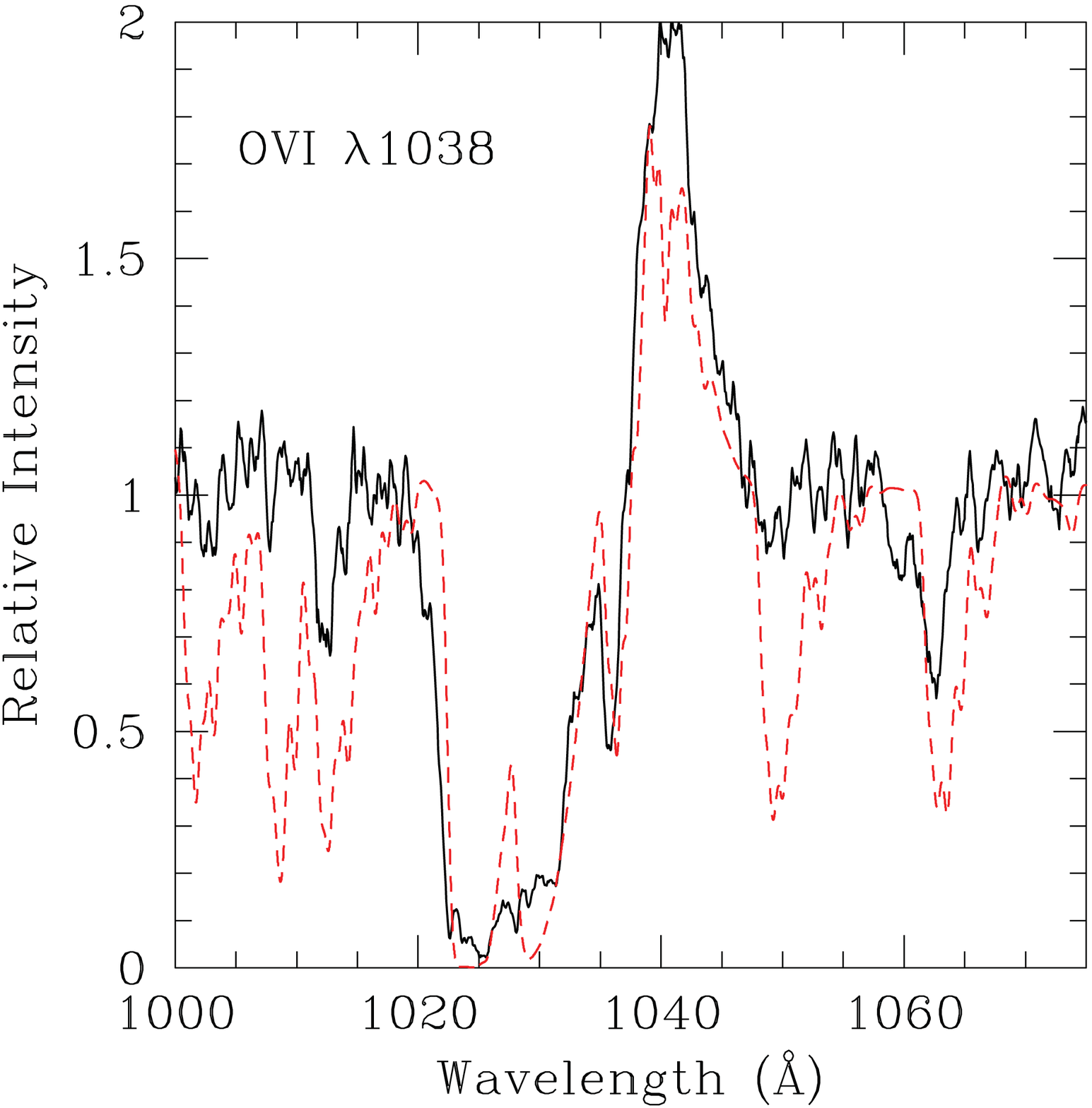}
\plotone{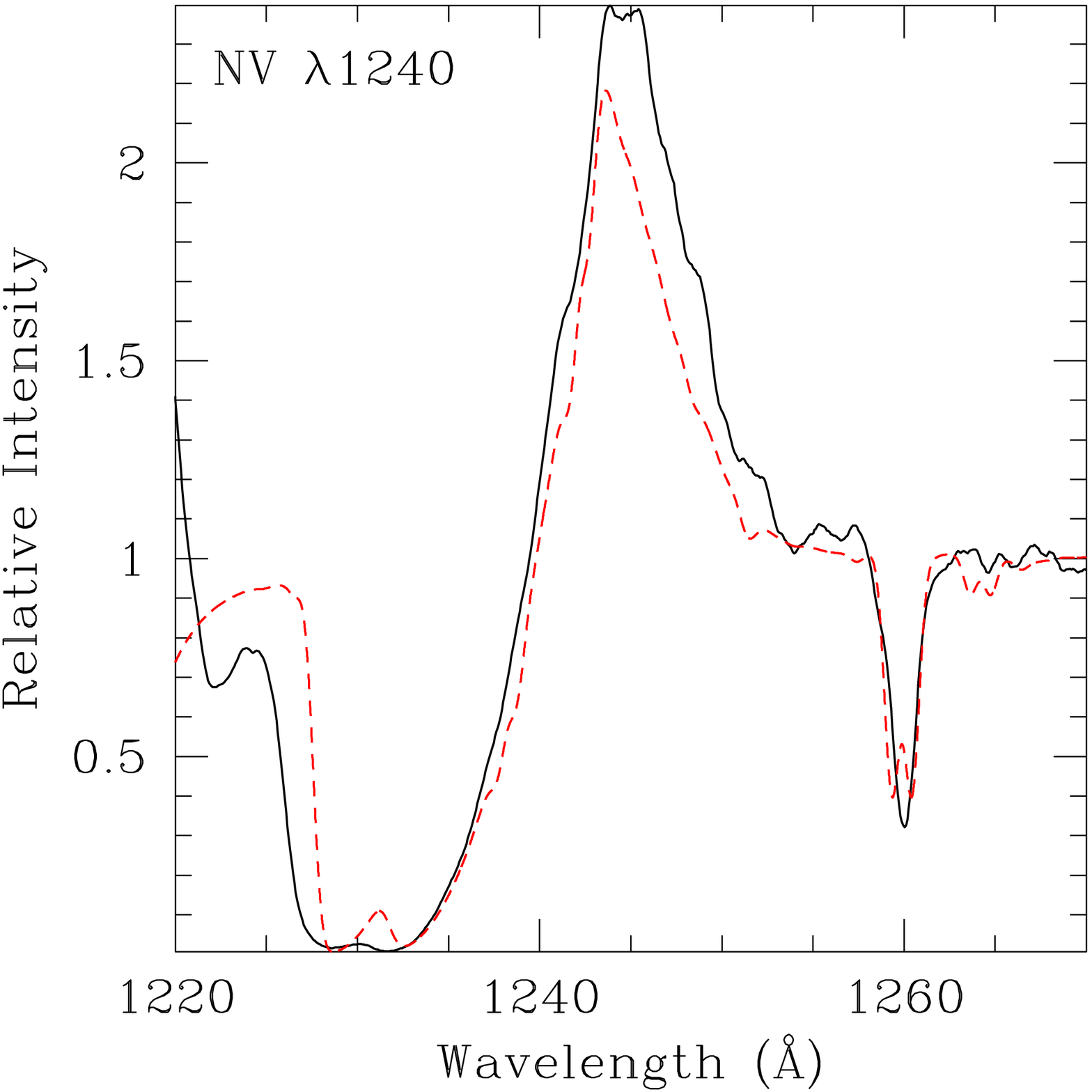}
\plotone{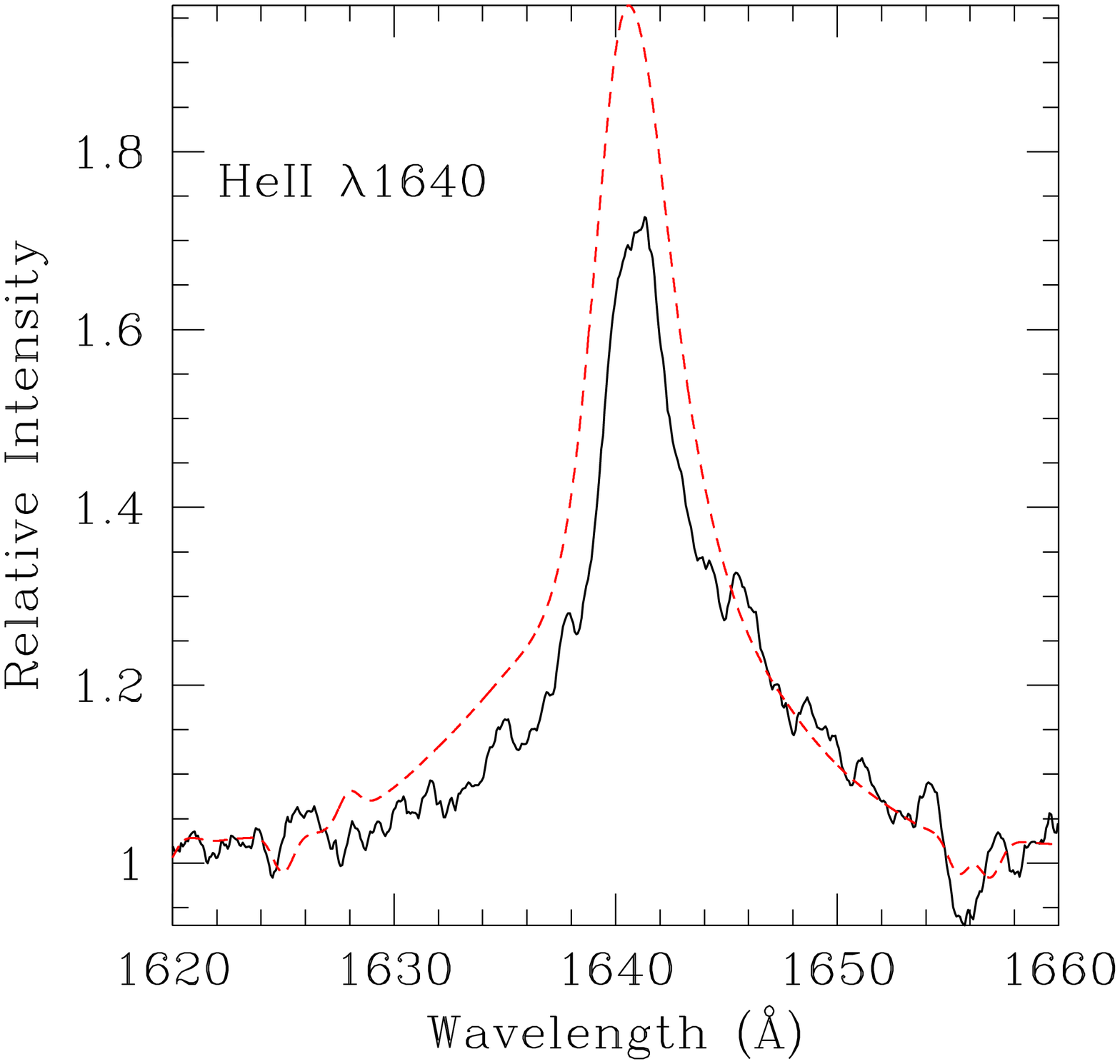}
\caption{\label{fig:LMC079-1_UV} Fit of the model (red dashed) to the spectrum of LMC079-1 in the UV for O\,{\sc vi} $\lambda$1038 (left), N\,{\sc v} $\lambda$1240 (middle), and He\,{\sc ii} $\lambda$1640 (right). }
\end{figure}

\begin{figure}
\epsscale{0.32}
\plotone{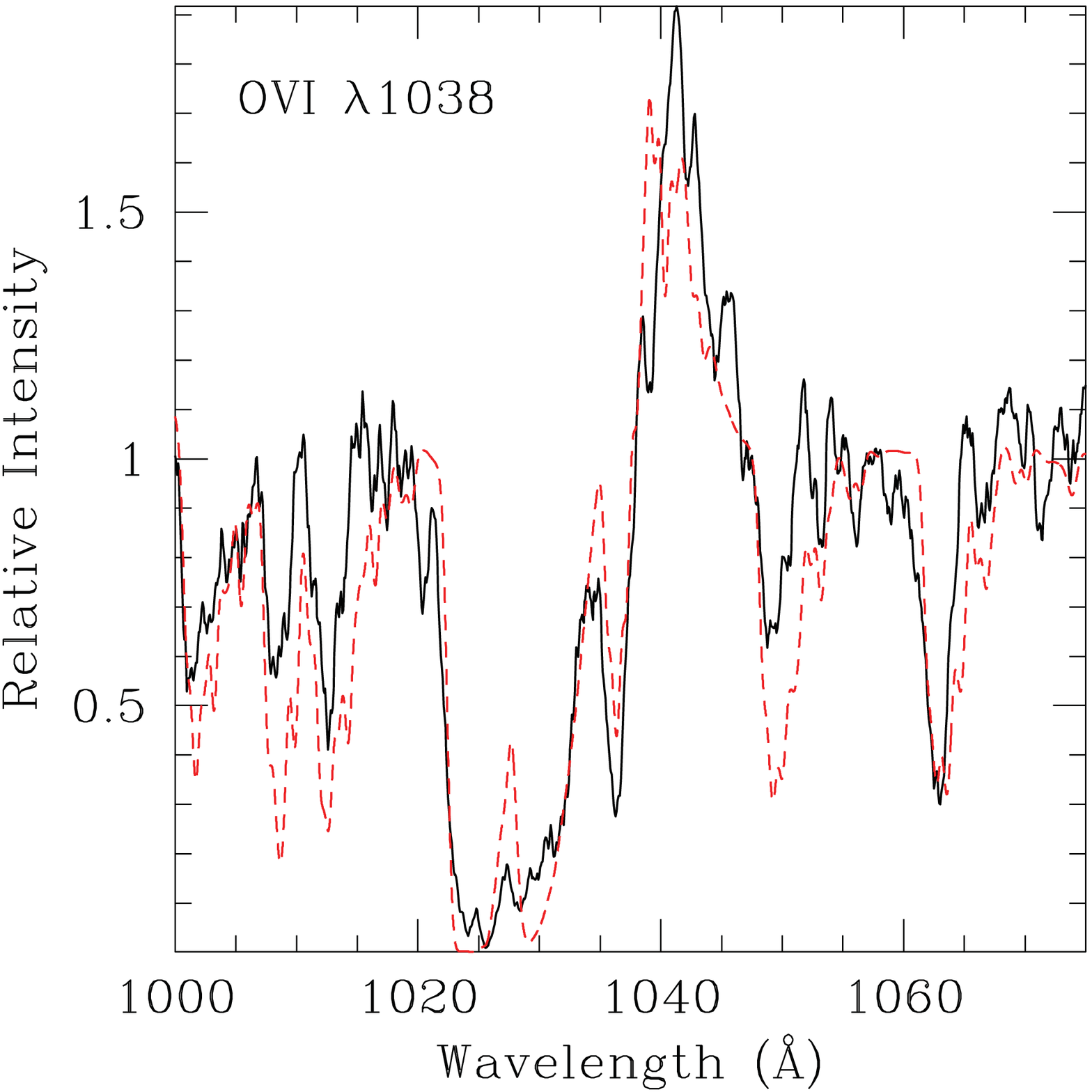}
\plotone{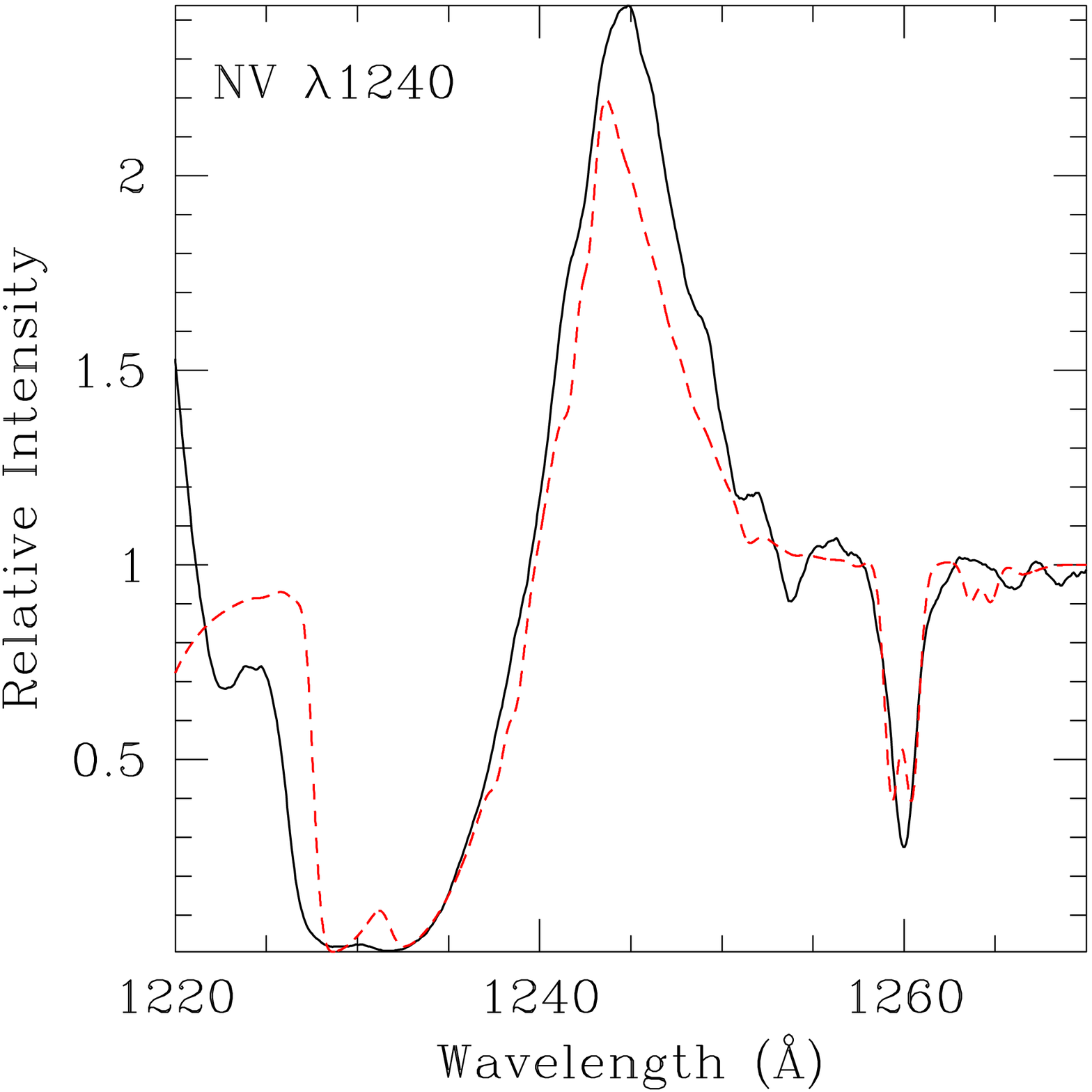}
\plotone{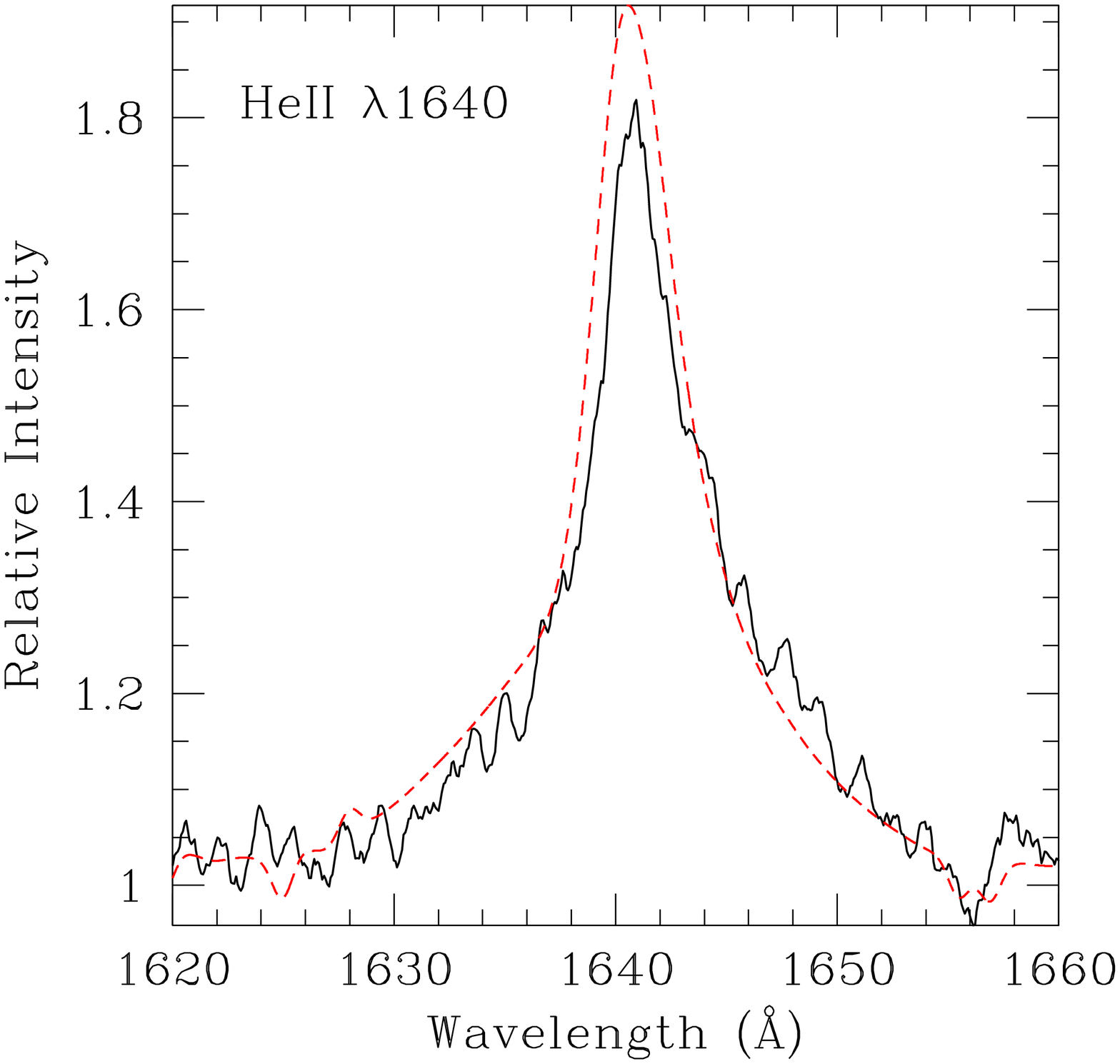}
\caption{\label{fig:LMC170-2_UV} Fit of the model (red dashed) to the spectrum of LMC170-2 in the UV for O\,{\sc vi} $\lambda$1038 (left), N\,{\sc v} $\lambda$1240 (middle), and He\,{\sc ii} $\lambda$1640 (right). }
\end{figure}

\begin{figure}
\epsscale{0.32}
\plotone{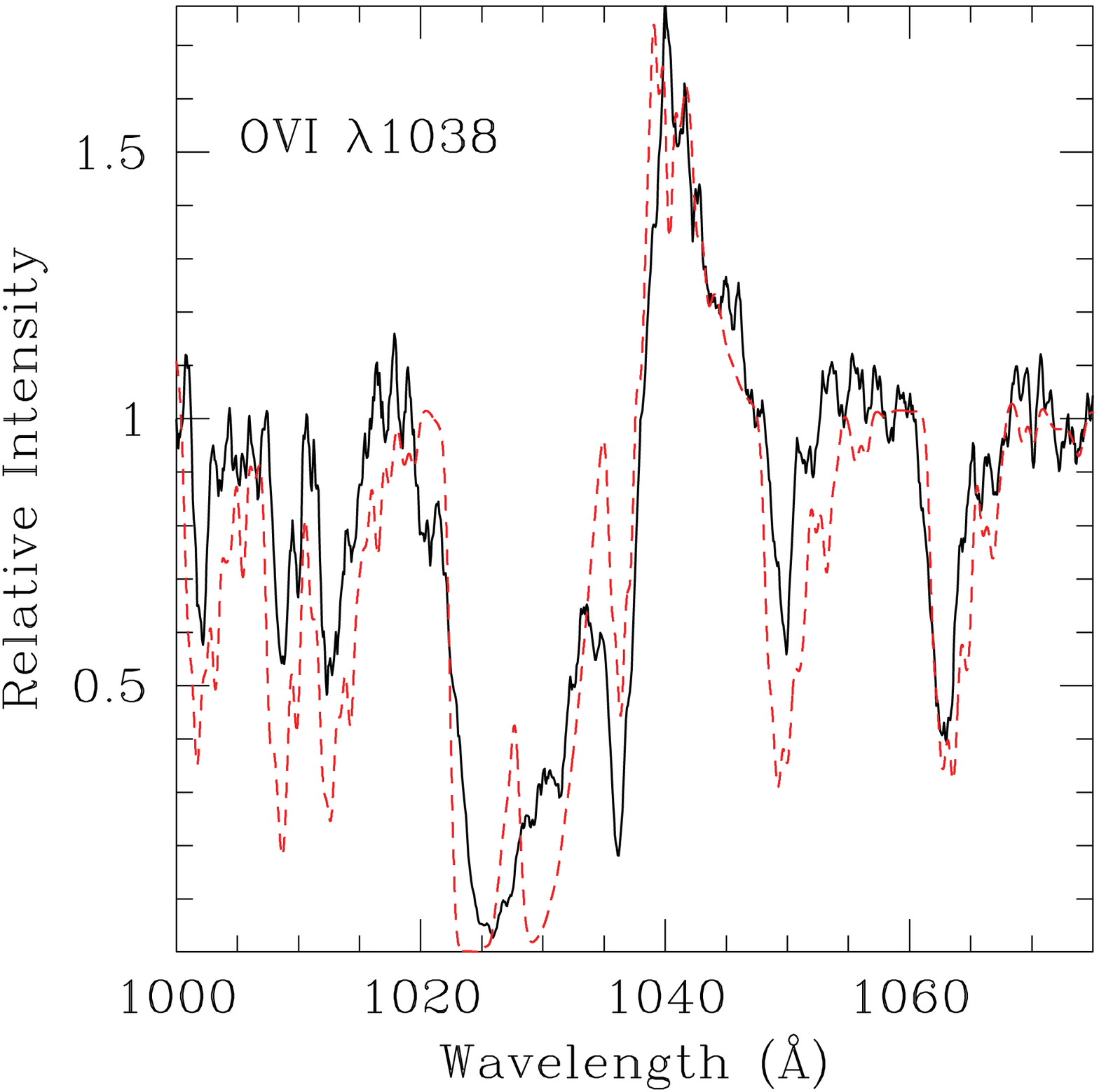}
\plotone{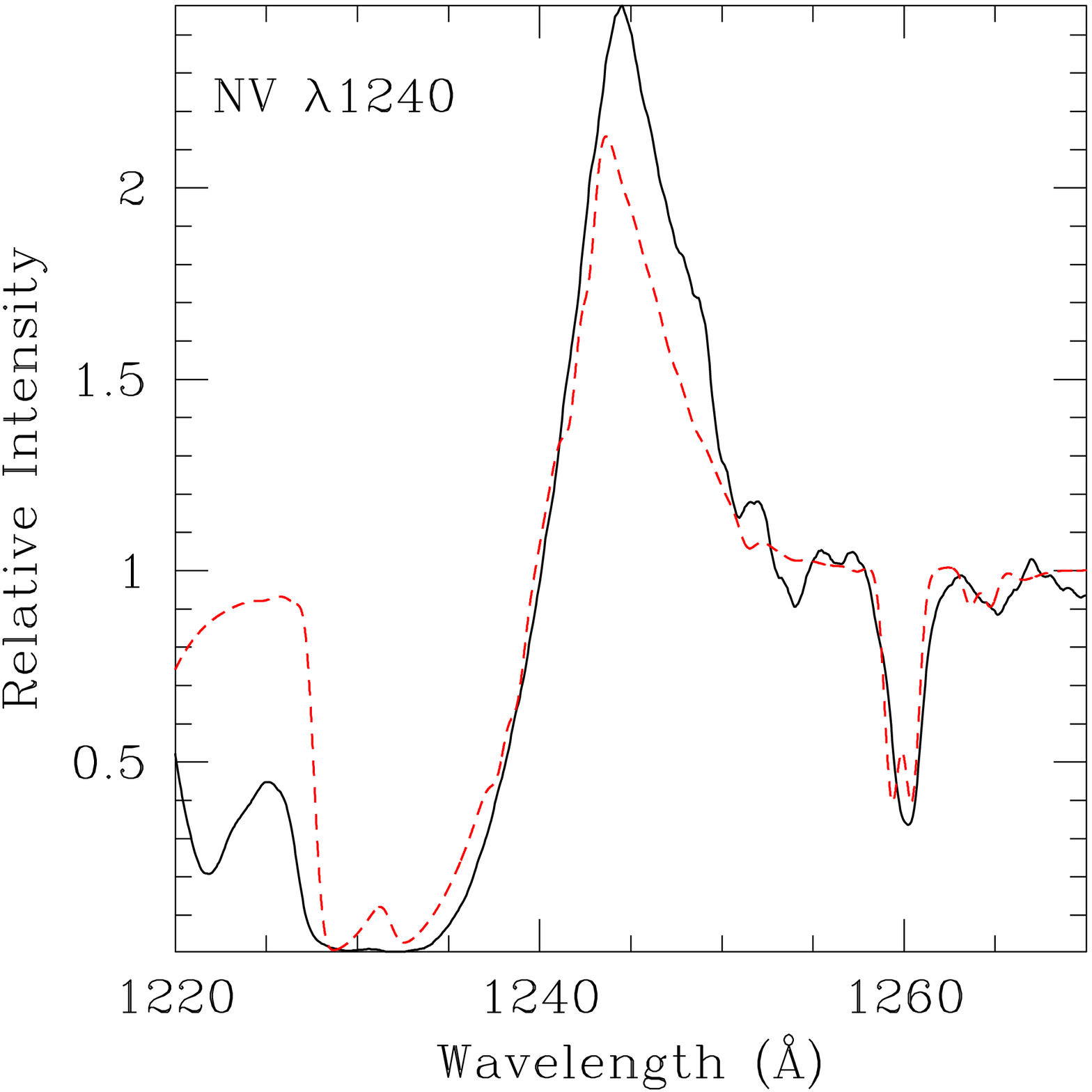}
\plotone{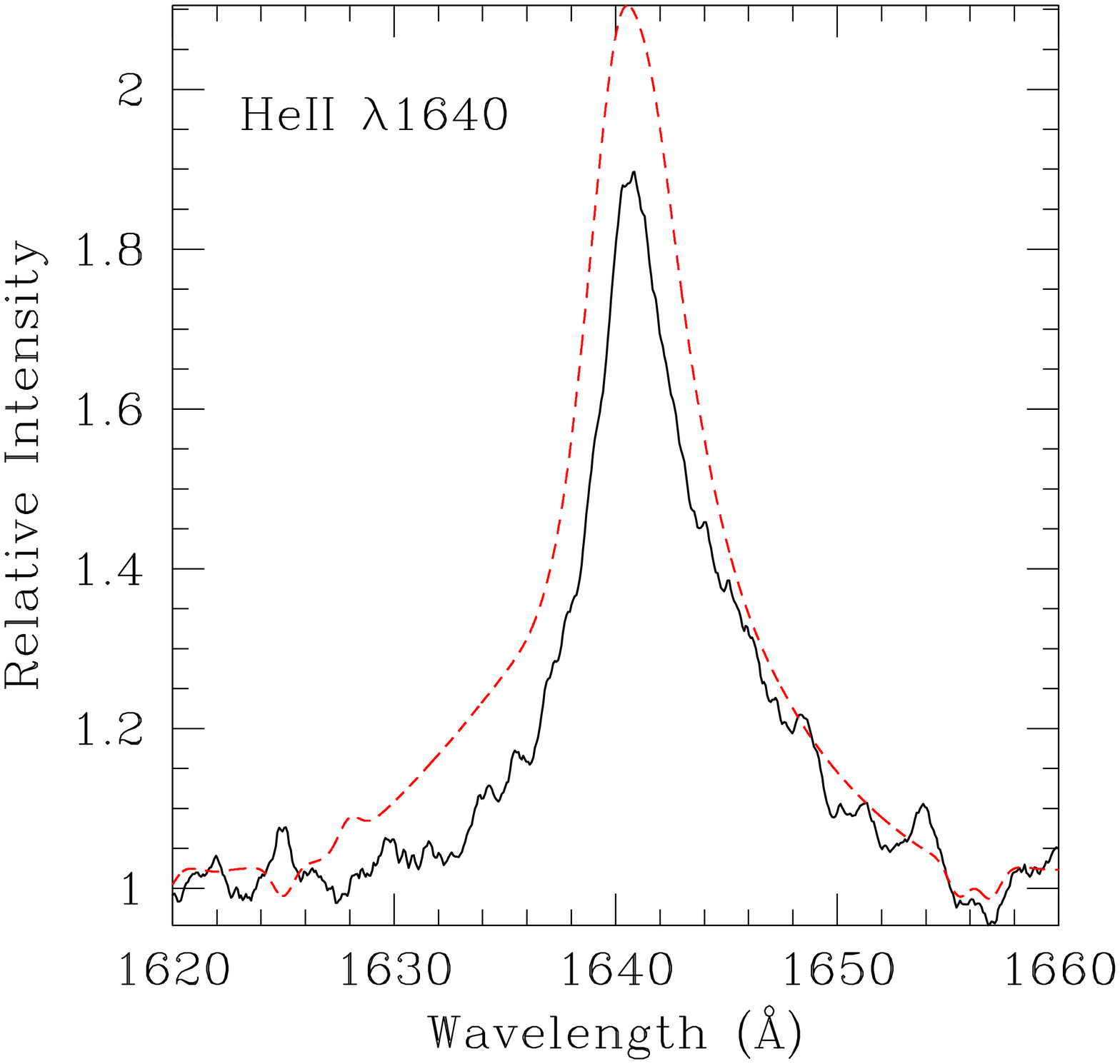}
\caption{\label{fig:LMC277-2_UV} Fit of the model (red dashed) to the spectrum of LMC277-2 in the UV for O\,{\sc vi} $\lambda$1038 (left), N\,{\sc v} $\lambda$1240 (middle), and He\,{\sc ii} $\lambda$1640 (right).}
\end{figure}

\begin{figure}
\epsscale{0.5}
\plotone{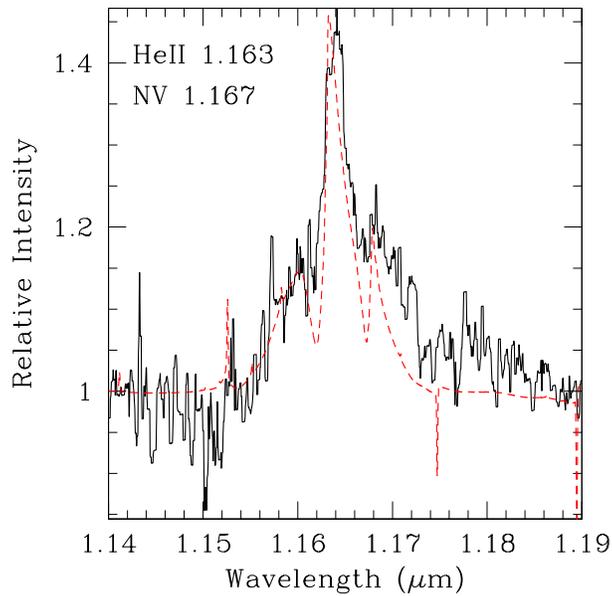}
\caption{\label{fig:LMC1702_NIR} Fit of the model (red dashed) to the spectrum (black) of LMC170-2 in the NIR for the He\,{\sc ii} $\lambda$11630, N\,{\sc v} $\lambda$11670 blend.}
\end{figure}

There was good agreement between the spectra and the model fit in the UV for LMC079-1, LMC170-2 and LMC277-2; the three stars with UV data. Figures~\ref{fig:LMC079-1_UV} - \ref{fig:LMC277-2_UV} show the fits to O\,{\sc vi} $\lambda$1038, N\,{\sc v} $\lambda$1240, and He\,{\sc ii} $\lambda$1640 for each of these three stars. Since we also had NIR spectra for LMC170-2, we show the fit to the He\,{\sc ii} $\lambda$11630, N\,{\sc v} $\lambda$11670 blend in Figure~\ref{fig:LMC1702_NIR}.

\begin{figure}
\epsscale{1.0}
\plotone{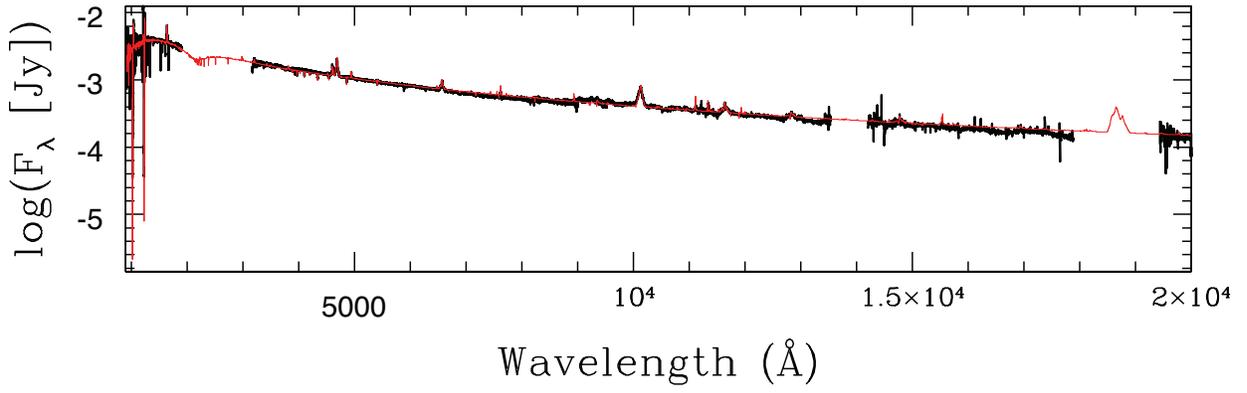}
\caption{\label{fig:sed} Model spectral energy distribution (SED) for LMC170-2 compared to the observed spectrum. The SED fits the UV, optical and NIR data well suggesting that the reddening has been well determined.}
\end{figure}

Additionally, the spectral energy distribution of the model almost perfectly matched our UV, optical and NIR spectra for LMC170-2, as is shown in Figure~\ref{fig:sed}. However, it should be noted that for stars with $T_{\rm eff} >$ 30,000~K, the slope spectral energy distributions (SEDs) will all look similar longwords of 1000 \AA\ as this is the Rayleigh-Jeans tail of the black body distribution.

The new physical properties, taking the UV and NIR data into account, are shown in Table~\ref{tab:allParams}. The uncertainty values for each fit are the same as those for LMC170-2 as shown in Table~\ref{tab:params}.

\section{Modeling Efforts: The Remaining Eight Stars}
\begin{figure}
\epsscale{1}
\plotone{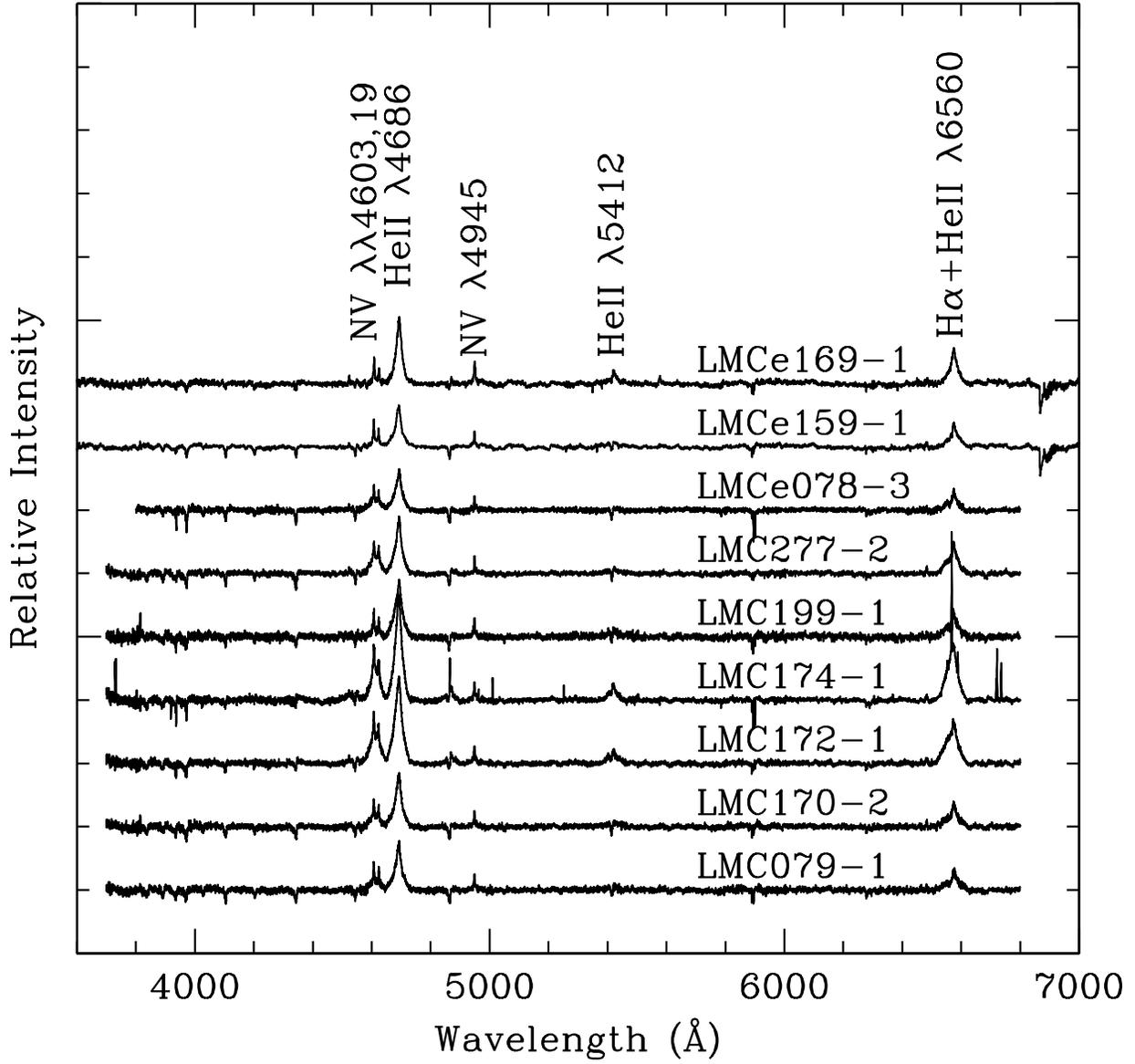}
\caption{\label{fig:o3s} Normalized spectra of all nine known WN3/O3s with prominent lines labeled. Notice the similarities between all nine spectra.}
\end{figure}

\begin{figure}
\epsscale{1.0}
\plotone{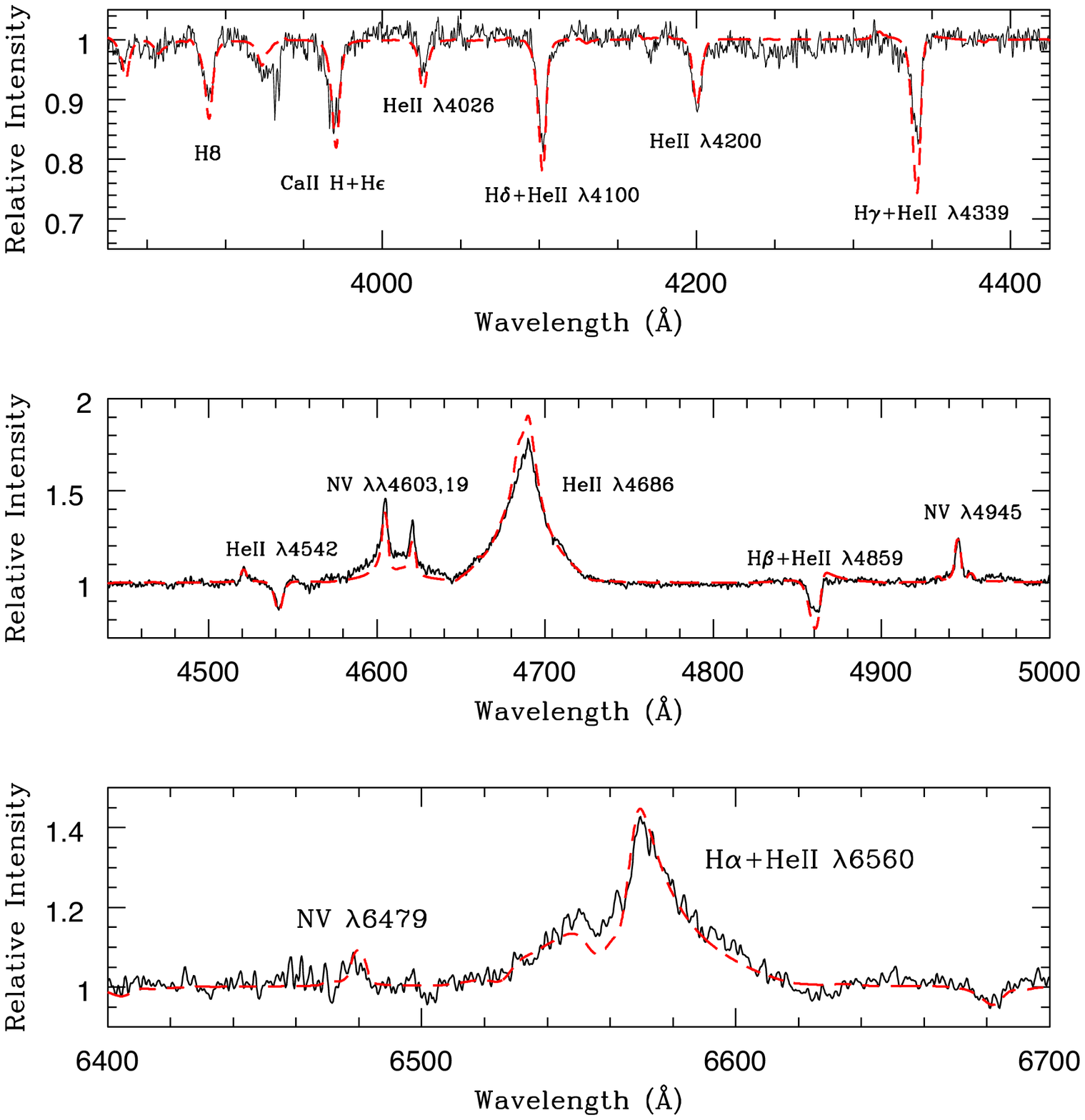}
\caption{\label{fig:LMC0791mod} Model fit (red dashed) to observed spectrum (black) for LMC079-1.}
\end{figure}

Figure~\ref{fig:o3s} shows the similarities between the spectra of all nine stars suggesting that our fit to LMC170-2 could be used as a starting point to fit the remaining eight stars, and that the changes in the individual parameters should be small\footnote{We are making the normalized optical spectra from this Figure available as fits files as part of this article.}. In fact, plotting the original LMC170-2 model with each of the eight spectra revealed that each spectrum matched the model to within 10\% and most to within 5\%. However, to obtain a parameter range for the WN3/O3s we still needed to model each star individually. The adopted models for all nine stars are shown in Figures~\ref{fig:LMC0791mod} - \ref{fig:LMCe1691mod}. For the most part we were able to obtain a good model fit to the spectra. However there were a few cases where we had some difficulty. 

We had problems with some of the fits of the N\,{\sc v} lines.  There are three different mechanisms forming N\,{\sc v} lines, and because of this, it was hard to get perfect fits to all of them at the same time. The first type is the N\,{\sc v} UV doublet produced through resonance scattering, and to a lesser extent, collisional excitation. In many of the stars, the P Cygni absorption and emission components have similar strength. The second is the N\,{\sc v} $\lambda \lambda$ 4603,4619 doublet arising from the 3s $^2$S to 3p $^2$P$_{\rm{o}}$ transition. This doublet has a dual nature, with a narrow photospheric component, and a broad wind component. The broad wind component is produced by recombination and is sensitive to the mass-loss rates, the effective temperatures, and the nitrogen abundance. The third is the N\,{\sc v} recombination lines at $\lambda$4945. For such lines there is usually one strong component and weaker satellite lines (whose presence may be masked by the S/N of the data). These lines are normally produced by recombination. In the WN3/O3 stars, the lines are narrow (their observed FWHM is primarily set by the rotation rate of the star) and are formed in the photosphere. While we were generally able to fit N\,{\sc v} $\lambda$4945, the fits to the broad component of N\,{\sc v} $\lambda \lambda$ 4603,4619 were too weak.   

The second reoccurring problem deals with the He\,{\sc ii} lines, particularly the fit of He\,{\sc ii} $\lambda$4686. The strength of He\,{\sc ii} $\lambda$4686 is strongly affected by the optical depth in the He\,{\sc ii} Lyman lines. For example, in one model (with $T_{\rm eff} =$ 95,000~K) both He\,{\sc ii} $\lambda$4686 and the H$\alpha$ / He\,{\sc ii} blend were well matched. However, a reduction of $T_{\rm eff}$ to 90,000~K causes a dramatic increase in strength of He\,{\sc ii} $\lambda$4686 (and $\lambda$1640) while leaving the strength of the H$\alpha$/He\,{\sc ii} blend relatively unaffected. Changes in surface gravity also cause a change in strength of He\,{\sc ii} $\lambda$4686, since this affects the flux shortward of 2280\,AA, and hence the population of singly ionized He.

Finally, absorption lines of H and He\,{\sc ii} are readily seen. These lines are not purely photospheric since they are influenced by the adopted mass loss rate, velocity law, and filling factor. In addition to their strength their radial velocities are model dependent. 

Here we will briefly describe the model fits for each star and identify stars that suffer from the modeling challenges detailed above.\\

\textbf{LMC079-1:} Figure~\ref{fig:LMC0791mod}. For LMC079-1 we had both COS UV and MagE optical data. The blueward side of the UV PCygni lines could be better fit with a terminal velocity larger than 2600 km s$^{-1}$ but to keep everything consistent, we adopted one terminal velocity for all nine stars. The final model of LMC079-1 fits each of the lines well except for the He+H lines ($\lambda$4340, $\lambda$4861, etc.) for the reasons discussed above.\\

\textbf{LMC172-1:} Figure~\ref{fig:LMC1721mod}. This model again suffers from the He+H line issue. Neither H$\alpha$ nor He\,{\sc II} $\lambda$4686 are particularly well fit (H$\alpha$ too weak while He\,{\sc II} $\lambda$4686 too strong), but since they are both highly mass-loss dependent, a model was adopted that was a compromise between the two of them.\\

\textbf{LMC174-1:} Figure~\ref{fig:LMC1741mod}. The spectrum of LMC174-1 is noisier than the rest of the WN3/O3 spectra. Again, a compromise has been made in the mass-loss rate to better fit both He\,{\sc II} $\lambda$4686 (too strong) and H$\alpha$ (too weak).

\textbf{LMC199-1:} Figure~\ref{fig:LMC1991mod}. Again, the He+H lines could be better matched. However, the nitrogen doublet is practically perfect which is rare given the three different ``types" of N\,{\sc v}, as discussed above. While $\beta$ could be increased a tad to better fit the profile of H$\alpha$, we are keeping $\beta$ constant at 1 to allow for better comparisons between the physical parameters.\\

\textbf{LMC277-2:} Figure~\ref{fig:LMC2772mod}. This model does not display the He+H line issue and instead manages to match most lines quite well. The one exception is He\,{\sc II} $\lambda$4686 which is again a compromise between an over fit He\,{\sc II} $\lambda$4686 and under fit H$\alpha$.\\

\textbf{LMCe078-3:} Figure~\ref{fig:LMCe0783mod}. This model also matches the spectrum well with the exception of H$\alpha$. Again, a larger value for $\beta$ would better match the spectrum.\\

\textbf{LMCe159-1:} Figure~\ref{fig:LMCe1591mod}. Again, $\beta$ could be increased slightly to make the model match the spectrum a bit better, but it is a good match other than that and a slight issue with the He+H lines.\\

\textbf{LMCe169-1:} Figure~\ref{fig:LMCe1691mod}. This star again has a pretty noisy spectrum and was thus difficult to model. A compromise has been made between H$\alpha$ and He\,{\sc II} $\lambda$4686 though it also appears that $\beta$ could be a slightly lower value. While this star also suffers from the He+H issue, it is also possible that this issue is more due to a less-than-perfect spectrum.

As is shown and discussed above, there are a few reoccurring issues with each of the nine fits. In half of the stars a compromise between He\,{\sc II} $\lambda$4686 and H$\alpha$ was made to fit these two mass-loss sensitive lines where the model fit to He\,{\sc II} $\lambda$4686 was always too strong while the model fit to H$\alpha$ was always too weak. Additionally, in half of these stars, $\beta$ could have been raised or lowered slightly to better fit the profile of H$\alpha$ (increasing $\beta$ ``fills in" the profile). However, we held $\beta$ steady at 1 so we could better compare the other stellar wind parameters. In addition, there are the nitrogen and He\,{\sc ii} absorption line issues discussed above.

\subsection{Physical Properties of WN3/O3s}
\begin{figure}
\epsscale{1.0}
\plotone{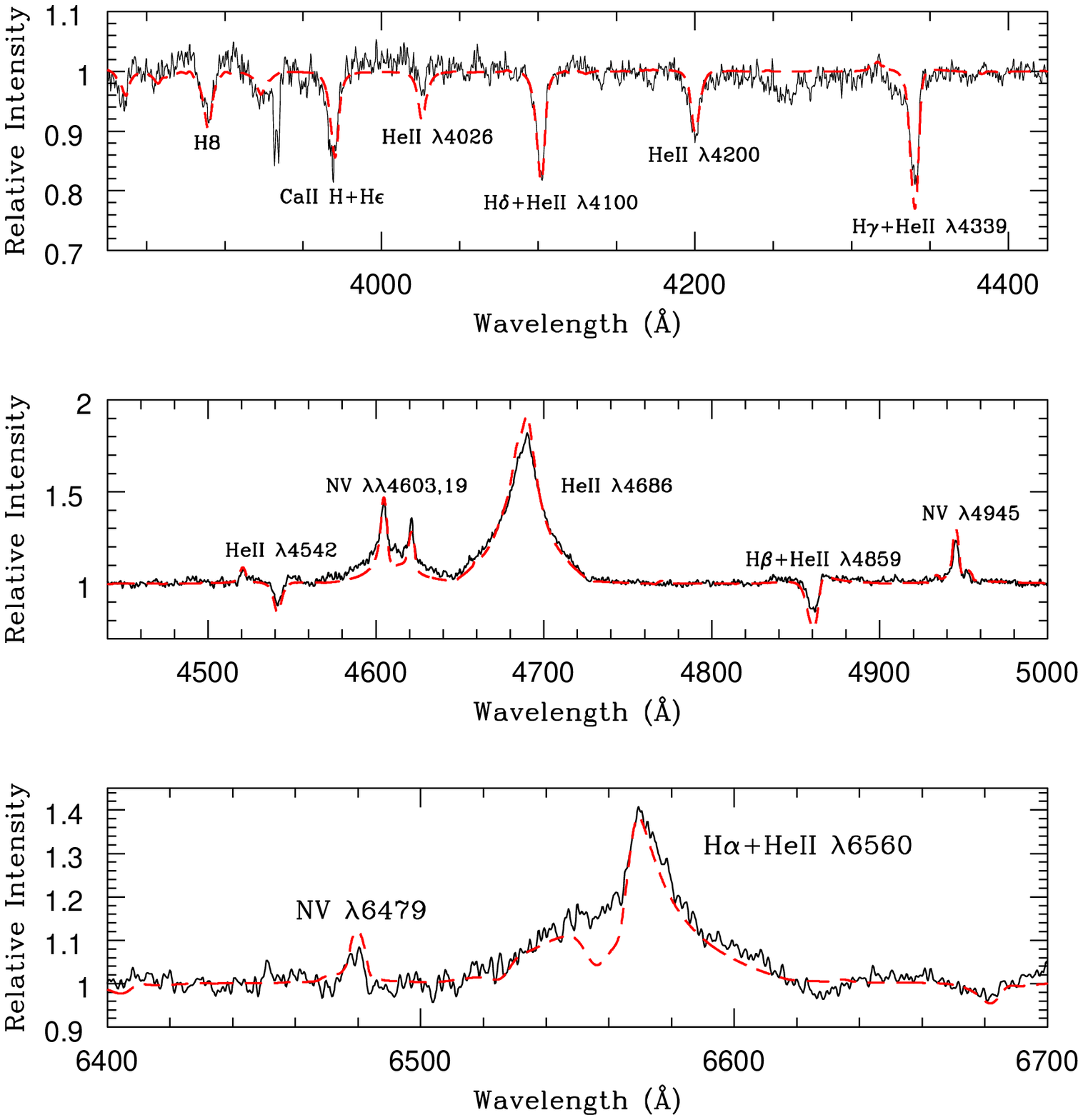}
\caption{\label{fig:LMC1702mod} Model fit (red dashed) to observed spectrum (black) for LMC170-2.}
\end{figure}

\begin{figure}
\epsscale{1.0}
\plotone{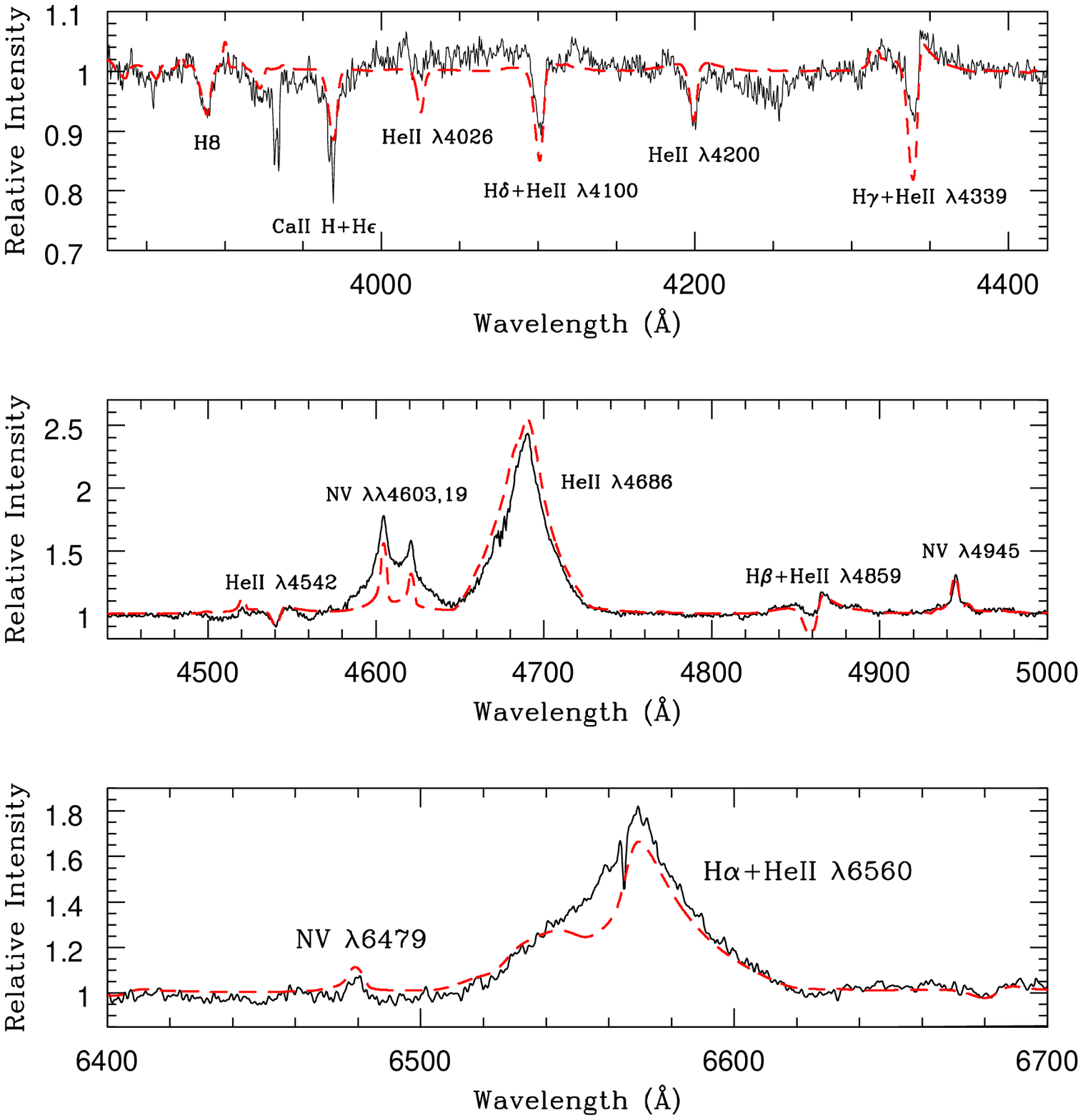}
\caption{\label{fig:LMC1721mod} Model fit (red dashed) to observed spectrum (black) for LMC172-1.}
\end{figure}

\begin{figure}
\epsscale{1.0}
\plotone{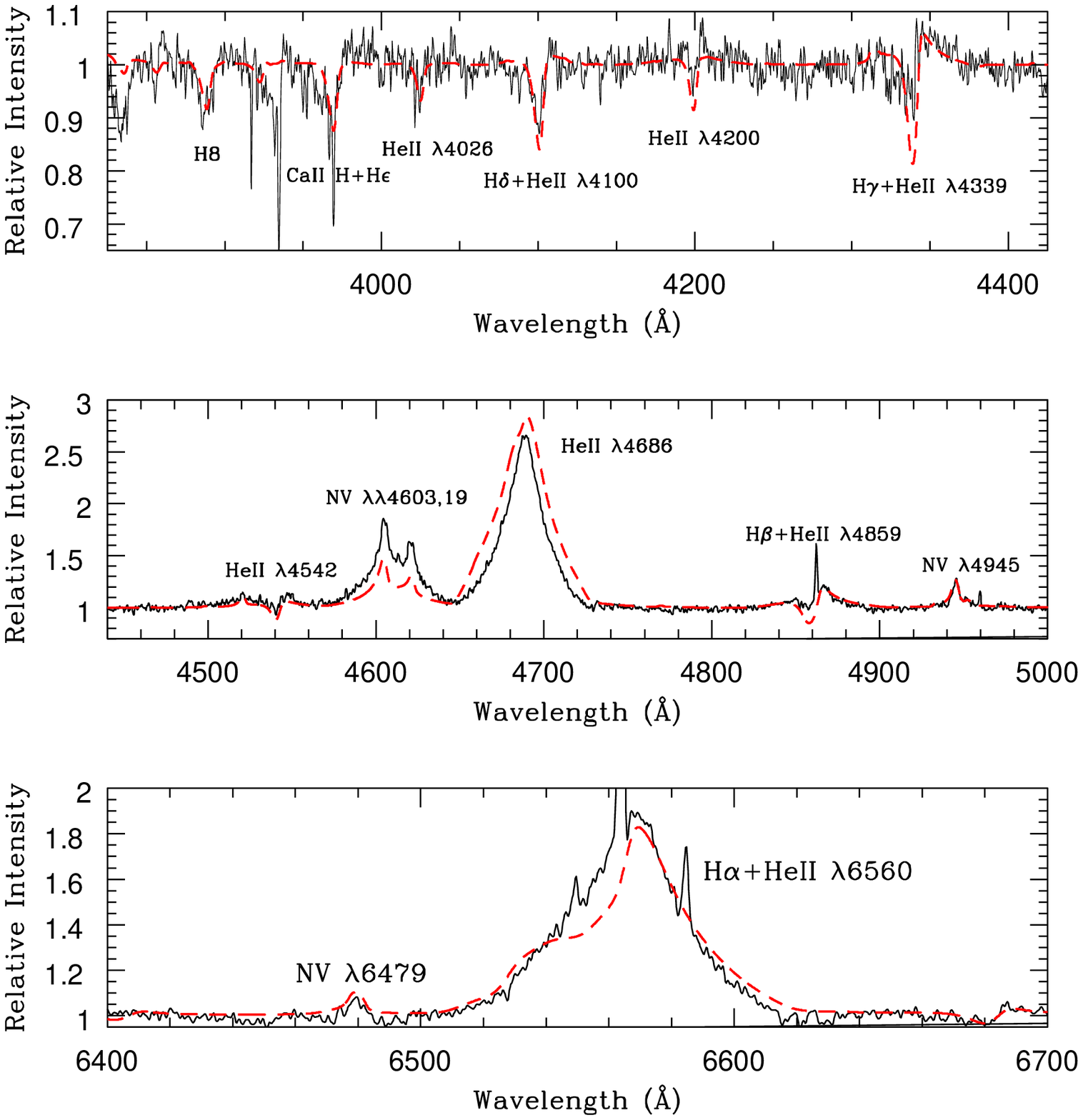}
\caption{\label{fig:LMC1741mod} Model fit (red dashed) to observed spectrum (black) for LMC174-1.}
\end{figure}

\begin{figure}
\epsscale{1.0}
\plotone{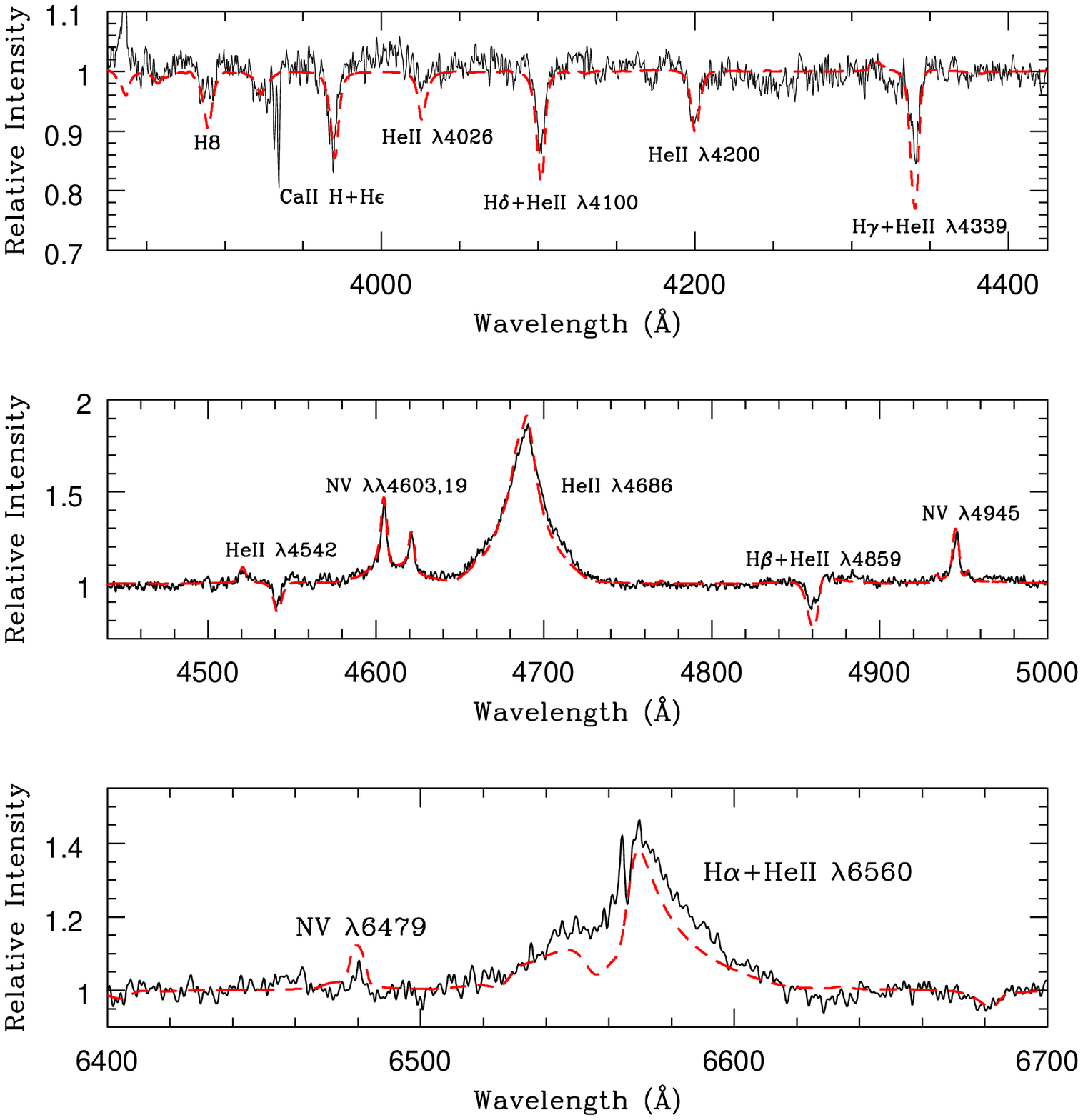}
\caption{\label{fig:LMC1991mod} Model fit (red dashed) to observed spectrum (black) for LMC199-1.}
\end{figure}

\begin{figure}
\epsscale{1.0}
\plotone{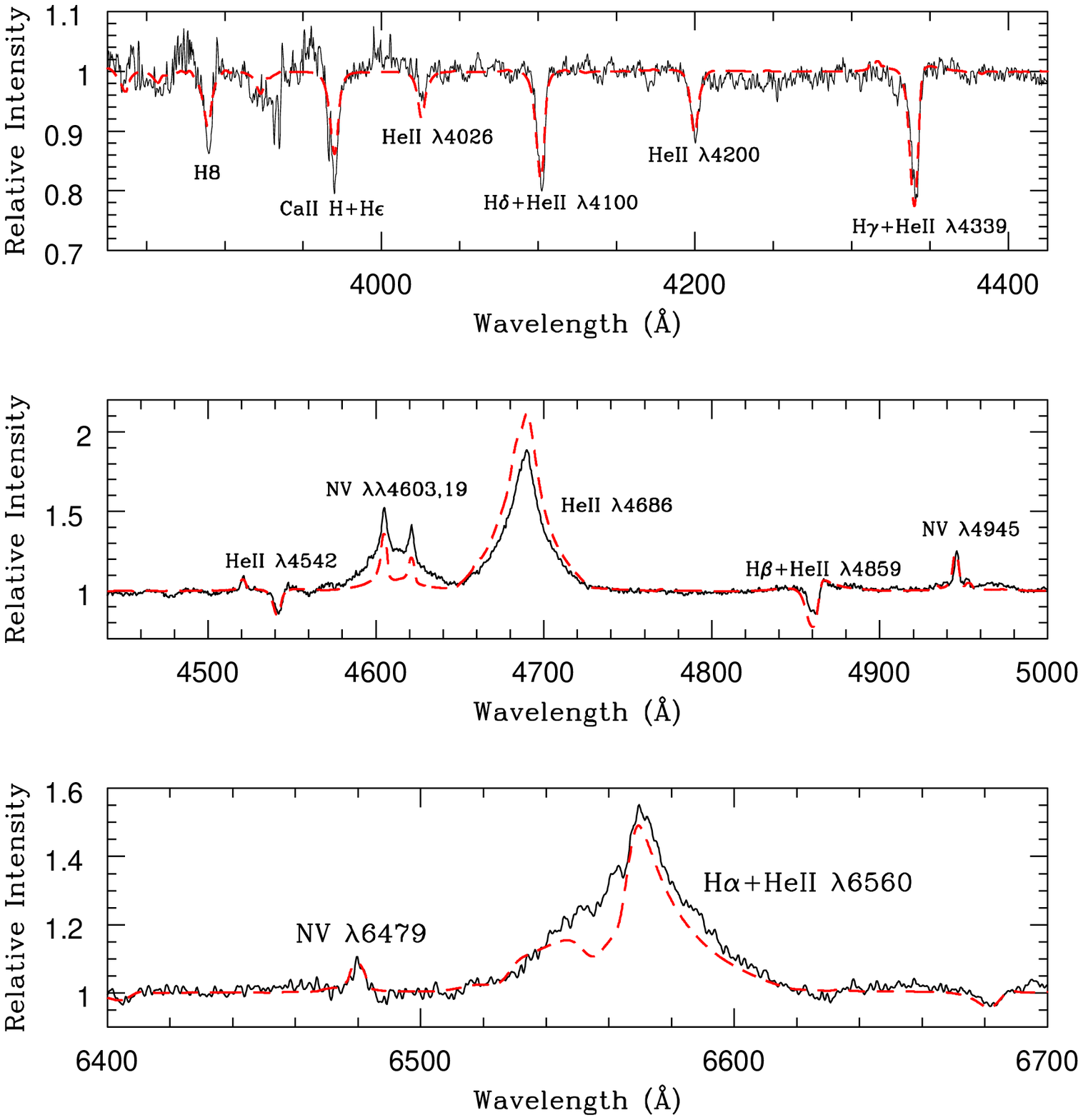}
\caption{\label{fig:LMC2772mod} Model fit (red dashed) to observed spectrum (black) for LMC277-2.}
\end{figure}

\begin{figure}
\epsscale{1.0}
\plotone{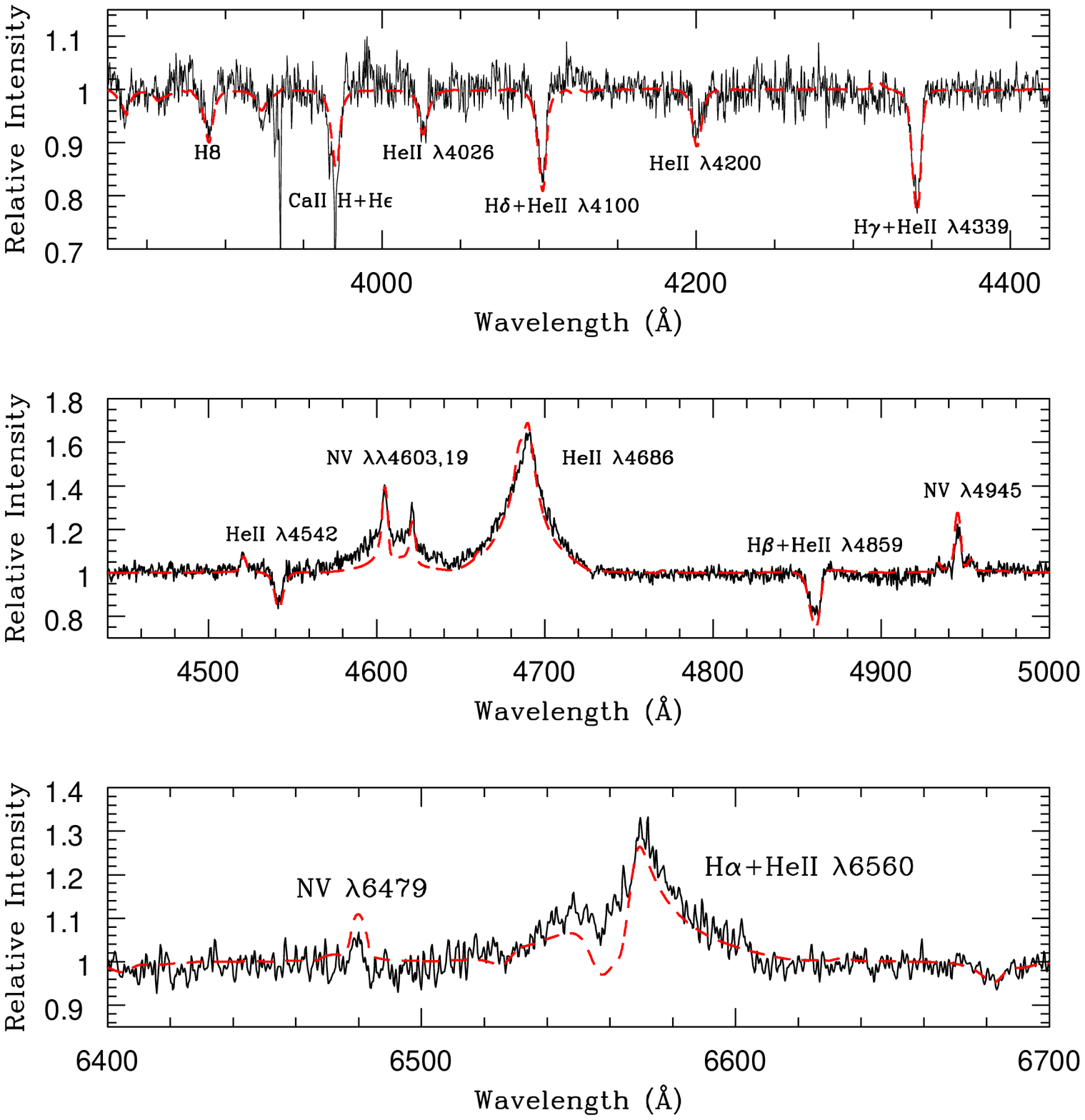}
\caption{\label{fig:LMCe0783mod} Model fit (red dashed) to observed spectrum (black) for LMCe078-3.}
\end{figure}

\begin{figure}
\epsscale{1.0}
\plotone{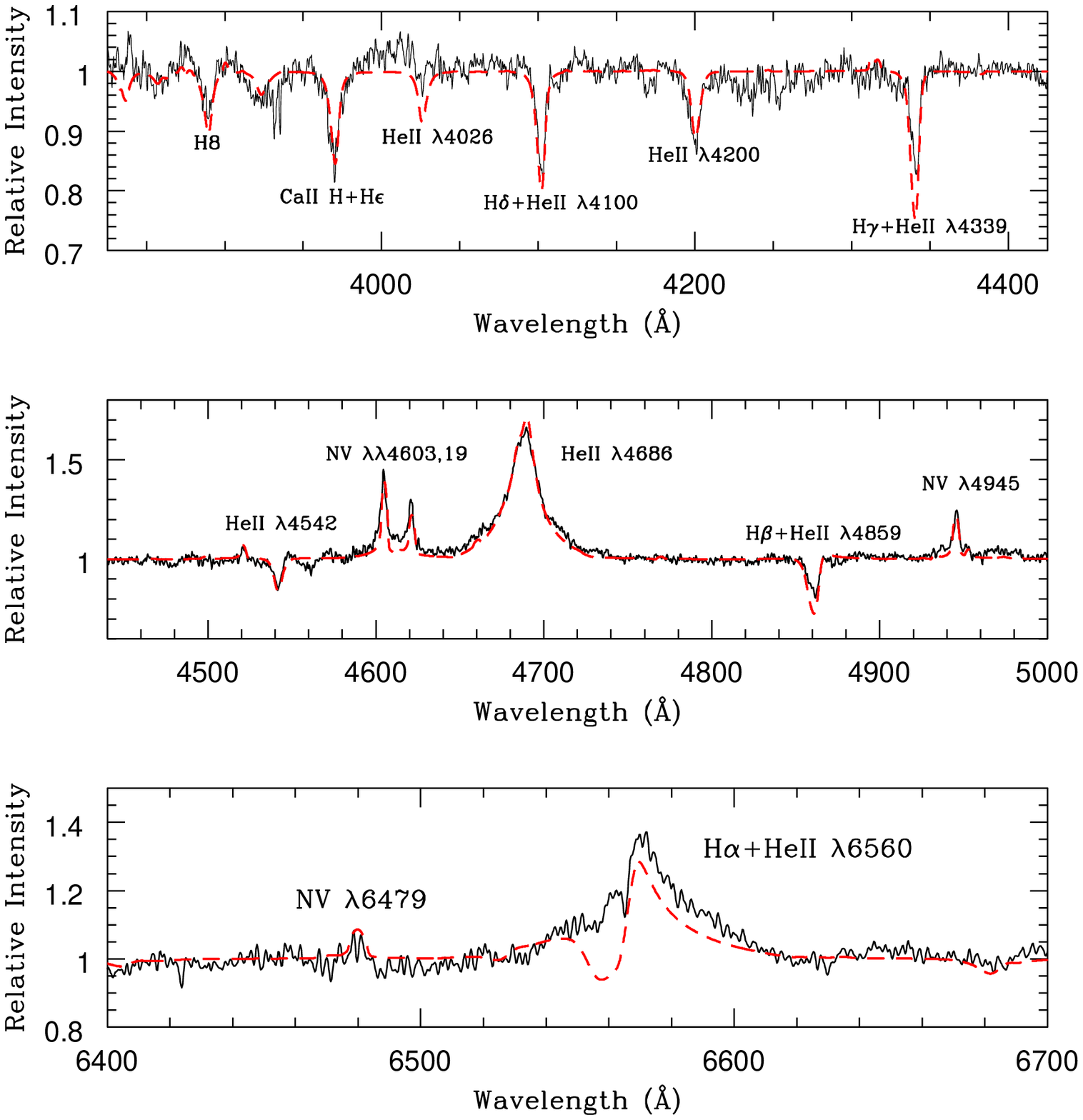}
\caption{\label{fig:LMCe1591mod} Model fit (red dashed) to observed spectrum (black) for LMCe159-1.}
\end{figure}

\begin{figure}
\epsscale{1.0}
\plotone{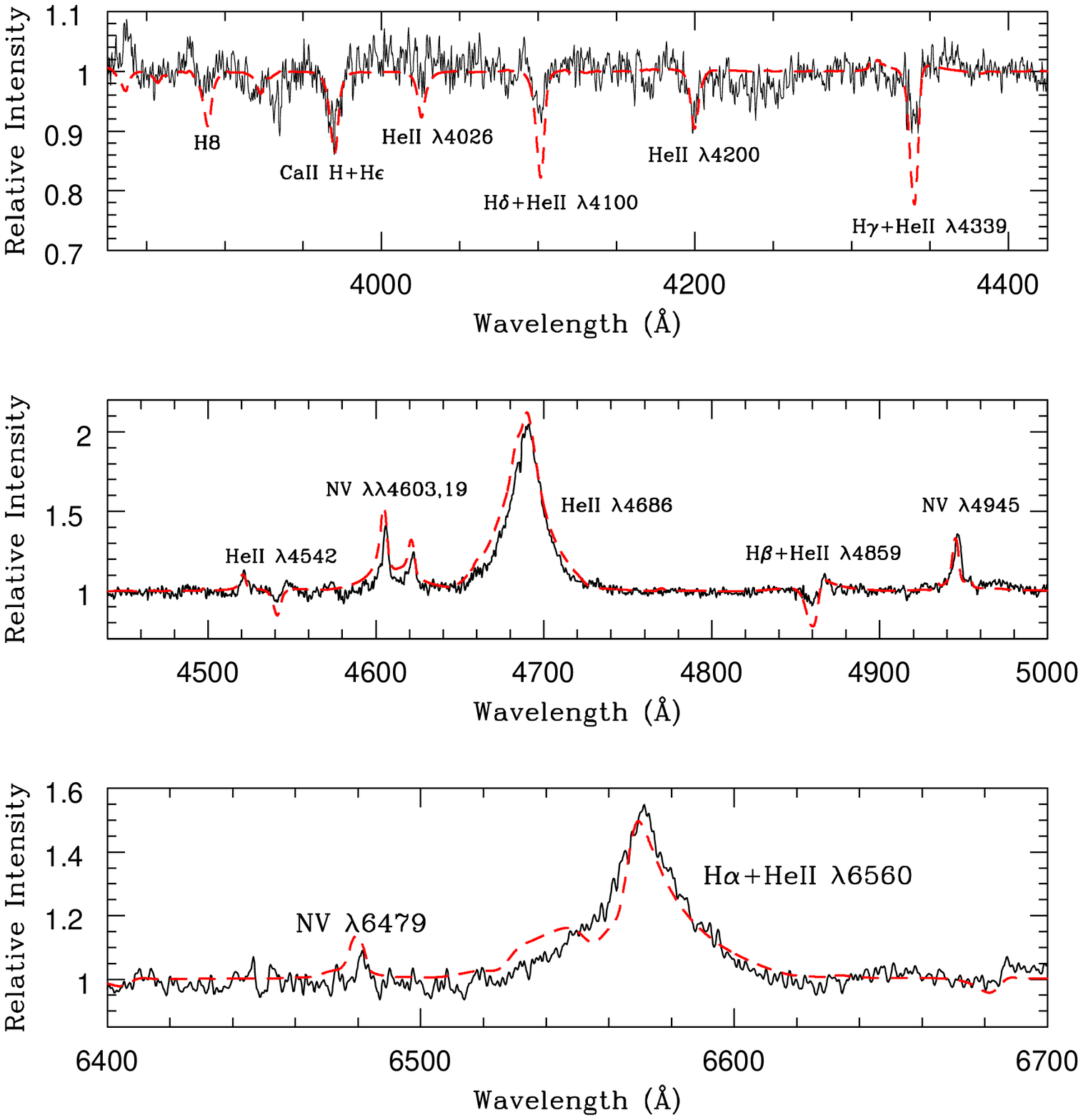}
\caption{\label{fig:LMCe1691mod} Model fit (red dashed) to observed spectrum (black) for LMCe169-1.}
\end{figure}

The physical properties of all nine stars are listed in Table~\ref{tab:allParams}. These are the parameters used to model the fits as shown in Figures~\ref{fig:LMC0791mod} - \ref{fig:LMCe1691mod}. Based on these values we determined the characteristic parameter range for the WN3/O3s. While most values were held constant (as is noted in Table~\ref{tab:allParams}), the temperatures varied between 95,000-105,000 K with the average temperature around 100,000K, the mass-loss rate varied by $\sim 40\%$, the helium to hydrogen ratio by around a factor of two and the nitrogen abundance varied between 0.011 and one half that (0.0055) by mass. So, overall, only slight variations in all but the mass-loss rate. 

As the temperature decreases, it becomes increasingly difficult to fit the strengths of the CNO lines. C\,{\sc iv} $\lambda$1550, a line not present in our spectra, appears in models with effective temperatures lower than 80,000 K. While we could have fixed this by setting the carbon abundance to zero, it is not physically realistic for a WN star to completely lack carbon (see below). Similarly, models with lower effective temperatures (70 K and 80 K) contain N\,{\sc iv} $\lambda$4058, which is not present in our spectra. Attempting to fit this lack of N\,{\sc iv} with a lower-temperature model would require changing the N abundance to an unrealistic value. Also, notice the lack of He\,{\sc i} $\lambda$4471 in the model fits as shown in Figures~\ref{fig:LMC1702mod} - \ref{fig:LMCe1691mod}. This also suggests a high temperature. Finally, stars with lower effective temperatures (and luminosities) have difficulties driving the observed mass-loss rates. Thus, we concluded that a model in a lower temperature regime could not be used to fit these stars. As the temperature increases above 110,000 K, both He\,{\sc ii} $\lambda$4200 and O\,{\sc vi} $\lambda$1038 become too weak, putting an upper limit on the temperature. The overall temperature range is therefore somewhere between 80,000 - 110,000K. 

During the hydrogen burning CNO cycle in massive stars, the initial carbon and oxygen are converted into nitrogen. Within the core, the abundances of these elements quickly reach equilibrium values. Mixing brings the core equilibrium products to the surface. The fact that we needed greatly enhanced nitrogen in order to model these stars suggests we can simply adopt the CNO equilibrium values for oxygen and carbon. Though, it should be noted that we cannot put a good constraint on the carbon abundance. (The values for oxygen and carbon are so decreased that Hainich et al.\ 2014 used the approximation that the oxygen abundance is zero.) The sum of the mass fractions of C, N, and O will remain essentially constant as C and O are converted into N.  The sum of these are essentially equivalent to the metallicity, {\it z}, as other metals are trace elements. For the LMC we expect that the metallicity is approximately half solar, consistent with analysis of H\,{\sc ii} regions and B-star abundances (see, e.g., Table 1 in Hainich et al. 2014).  However, what this actually means in terms of the mass fractions is a bit uncertain given recent revisions to the solar abundances. This is particularly true for oxygen (Asplund et al.\ 2009) and its change from the older values (e.g., Cox 2000). Traditionally, the metallicity adopted for the LMC is 0.008; with the Asplund et al.\ (2009) solar values it would be more like 0.006.

In modeling the stars, we adjusted the N abundances (in combination with other parameters) to obtain the best fit to the N\,{\sc v} lines. We used a mass fraction of either 0.011 or 0.0055 (roughly an equal number of times for each) depending on the strengths of these lines. Since these values are consistent with what we expect for the CNO equilibrium values for the LMC, we simply fixed the carbon and oxygen values at their approximate CNO equilibrium values, roughly half of what we expect for solar metallicity.  Thus we adopted mass fractions of 1.0E-4 for carbon, and 8.05E-5 for oxygen, guided by the evolutionary models of Ekstr\"{o}m et al. (2012). Our oxygen mass fraction may be a bit low given the poor fit to O\,{\sc v} $\lambda$1370, which shows a small ($\sim$5\%) P Cygni profile in the spectrum of LMC170-2. The expected CNO equilibrium abundance for N is 0.0065, a bit lower than what we typically used, but as Hainich et al.\ (2014) note, modelers of Galactic WN stars have usually had to adopt a mass fraction 2$\times$ higher than expected\footnote{In their modeling of LMC WRs, Hainich et al.\ (2014) adopted a surprisingly low mass fraction for nitrogen, 0.004.}. 

\begin{figure}
\epsscale{0.5}
\plotone{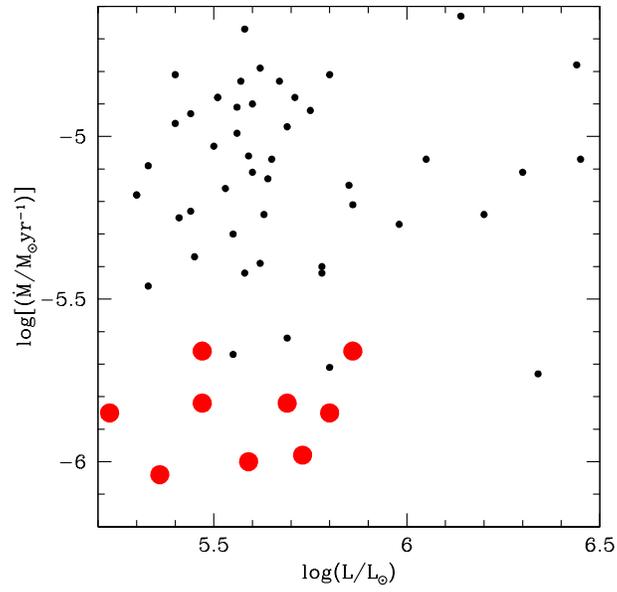}
\caption{\label{fig:massvary} A comparison of the luminosity vs.\ mass-loss rate for all nine WN3/O3s. The small black circles represent the early-type LMC WNs (WN3s and WN4s) analyzed by Hainich et al.\ (2014). The WN3/O3s are represented as large red circles.}
\end{figure}

For the most part, we discussed the stellar wind parameters such as our values for $\beta$ and the terminal velocities previously. The surface gravity is dependent on the effective temperature and $\frac{\dot{M}}{\sqrt{fR^3}}$. Since the temperatures and radii are all similar (our radii determination is discussed below), we adopt $\log g$ [cgs] $= 4.95$ dex for eight stars, consistent with the fits to the wings of H$\gamma$. For one star, LMC079-1, we adopt a slightly lower surface gravity of $\log g$ [cgs] $= 4.85$ dex given its slightly lower temperature and better fit to the wings of H$\gamma$. The mass-loss rates for all nine stars differ slightly from one another. This is shown visually in Figure~\ref{fig:massvary}. While the majority of the WN3/O3s have mass-loss rates much lower than those of ``normal" early-type WNs in the LMC (as modeled by Hainich et al. 2014), there are three ``normal" WNs that appear to fall within the same mass-loss regime as the WN3/O3s. However, all three stars are much brighter than the WN3/O3s (M$_V \sim -2.5$) with M$_V$ between -4 and -5.5 suggesting that they are not the same type of star. These three stars (BAT99-049, BAT99-072, and BAT99-025) have been classified by Foellmi et al (2003). BAT99-049 is a WN4:b+O8V binary system with a known period and has a fairly bright M$_V = -5.49$. However, BAT99-072 and BAT99-025 are both WN4 stars with absorption lines. While they do not appear to be related to WN3/O3s due to their magnitudes, we are still investigating whether they could somehow be connected. 

\subsection{Reddenings, Radii and Masses}
The presence of absorption lines in the spectra of our WN3/O3s provides a nearly unique opportunity: we can determine these stars' current masses, arguably their most fundamental physical property.  We do so as follows. As stated above, we have used the wings of H$\gamma$ to obtain values of the stars' surface gravities; these all had roughly the same value $\log g$ [cgs]$ = 4.85-4.95$ dex. For each star, we can then match the model's fluxes to the observed data, corrected for reddening.  This then allows us to scale the model's radius $R_*$ to that needed to produce the same flux.  Since $g/g_\odot = M/R_*^2$, we can then estimate the mass $M$, as $\log M =4.95-4.44+2\log{R_*}$, where we have adopted $\log{g_\odot}=4.44$. These values are shown in Table~\ref{tab:allParams}. Allowing for a 0.1 mag uncertainty in $V$, and a 5000 K uncertainty in the effective temperature, the uncertainty in the radius is around 6\%. Adopting this, and an error in log g of 0.05 dex, the uncertainty in the masses is 20\%. We include the model's luminosity in Table~\ref{tab:allParams} after adjustment for the stellar radius. The uncertainty in this is 0.1 dex. 

The models had all been run with a single radius that was a good match to the UV and NIR fluxes of LMC170-2; in order to determine the radius for the other stars, we compared the (reddened) model fluxes to the observed fluxes (As mentioned in Section 2.1, the optical fluxes have been adjusted to match the broad-band $V$ magnitudes in order to adjust for differential slit losses between the program star and standard star observations.) In reddening the models, we used a foreground correction E(B-V)=0.08 using the Cardelli et al.\ (1989) Galactic reddening law with $R_V=3.1$, and, for most stars, an additional E(B-V)=0.05 using the Howarth (1983) LMC reddening law, also with $R_V=3.1$. The total reddening of E(B-V)=0.13 is a good approximation for most of the lightly reddened early-type stars in the LMC as shown by Massey et al.\ (1995).  Massey et al.\ (2007) discusses this further, and separates the foreground and LMC-internal components using the Schlegel et al.\ (1998) Galactic reddening map. The exceptions to this procedure were LMC174-1 and LMCe078-3, whose broad-band photometry indicated higher reddening, and we adopted an LMC component of E(B-V)=0.25 and 0.13, respectively, in addition to the E(B-V)=0.08 Galactic component\footnote{We caution that one should not over-interpret the broadband photometry in Table 2. For instance, naively it would appear that LMC079-1 is somewhat fainter and more heavily reddened than LMC170-2. We find that LMC079-1 is about 0.3~mag fainter in the UV according to our COS SED. This would be consistent with it being the more highly reddened object were it not for the fact that the shapes of the UV SED are identical for the two stars, which is inconsistent with there being a difference in reddening. For both facts to be true, the reddening law would have to be anomalous for one of these objects.  While we can not rule this out {\it a priori}, a more pedestrian explanation is that photometry in the crowded Magellanic Clouds is not precise.  The formal error on $B-V$ is typically 0.07, so the difference in $B-V$ colors of 0.08 is well within the combined errors.  Furthermore, if we had chosen to use the photometry of Massey (2002) instead, we would conclude that LMC079-1 is actually redder ($\Delta(B-V)=+0.28$) than LMC170-2 rather than less red ($\Delta(B-V)=-0.07$), with nearly identical $V$ magnitudes.  Thus in correcting for reddening we have chosen to ``broad-brush" the photometry, and adopt an average reddening for all but two of the stars.}.

An example of how well our procedure worked is shown in Figure~\ref{fig:sed} for LMC170-2, the only star for which we have UV, optical, and NIR data. Note that the reddening laws are only defined down to around $\lambda 1250$ \AA\ so it is unsurprising that the shape of the reddened model does not perfectly match the shape of the observed data at shorter wavelengths.

\section{Discussion: Evolutionary Status}
\subsection{What We Know}

\begin{figure}
\epsscale{1}
\plotone{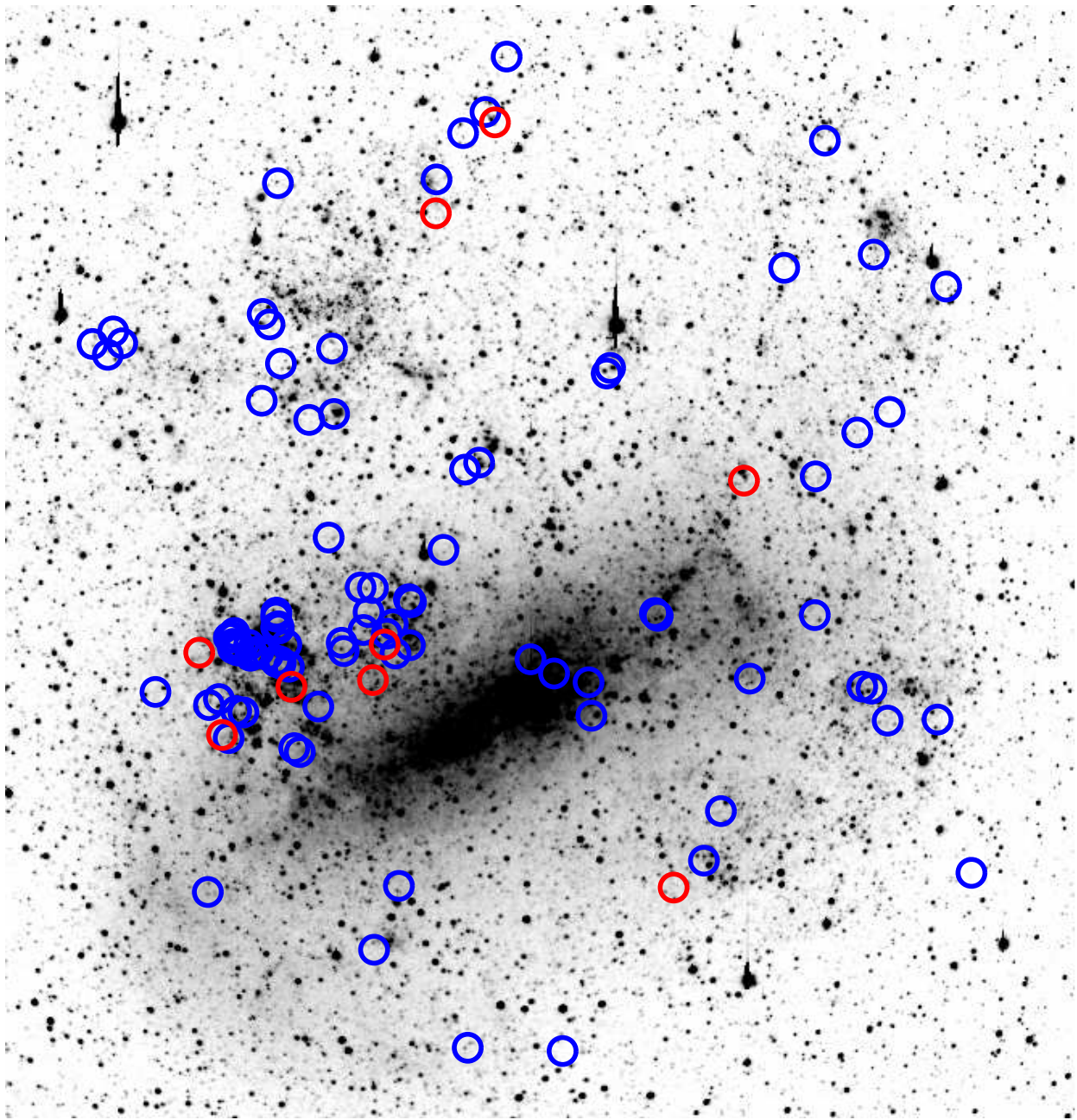}
\caption{\label{fig:locations} Locations of the nine known WO3/O3s (red) and known LMC WNs (blue). Note that the two distributions are roughly the same. There is a higher population of WRs near the LMC's star-forming OB associations.}
\end{figure}

The composition of the 9 analyzed WN3/O3 stars is in agreement with what we would expect for a WN-type LMC WR. For example, they have a CNO equilibrium abundance of nitrogen that is characteristic of the metallicity of the LMC; their mass fraction of around 0.008 proves that these stars didn't come from SMC-like metallicities. Additionally, as Figure~\ref{fig:locations} shows, the overall distribution of the WN3/O3s is similar to the distribution of LMC WNs with a higher population of both WNs and WN3/O3s near the LMC's star-forming OB associations. Figure~\ref{fig:locations} also shows that the WN3/O3s are spread throughout the LMC. If they were spatially separated, or more isolated from other LMC WRs, one might infer that the WN3/O3s were formed from different metallicity progenitors. However their spatial distribution suggests that this is not the case\footnote{In a recent preprint, Smith et al. (2017) claim that the WN3/O3s are more isolated from the O-type star population than are the other WRs. In order to test the isolation of the WN3/O3s further, we computed the projected separation between each WN3/O3 and the nearest normal WN2-4 star. The median value is $\sim$750\arcsec. This is essentially the same as the median separation of each WN2-4 from its nearest neighboring WN2-4 star, $\sim$730\arcsec. Thus we conclude that the WN3/O3s are no more ``isolated" than are the other early-type WNs, and that their progenitor populations have similar spatial distributions.}. Thus, their composition is most likely similar to other early-type WNs. 

These stars represent a significant fraction of the WN population of the LMC. Currently there are nine known or 6\% of the LMC WRs and 8\% of the LMC WNs. This suggests, as is discussed below, that they aren't formed from rare or improbable events.

As is shown in Figure~\ref{fig:o3s}, all nine of these stars have similar spectra. They all have the WR's characteristically broad emission lines with O3-like absorption lines. Their physical parameter range is quite small, as is shown in Table~\ref{tab:allParams}.

The absolute visual magnitudes for the WN3/O3s are quite faint and range between $M_V = -1.8$ to $-3.1$ with an average of $-2.6$. Hainich et al.\ (2014) finds an average $M_V = -3.8$ for ``normal" WN3s. Since the temperatures of the WN3/O3s are hotter than the WN3s, but their luminosities are comparable, these fainter magnitudes make sense. It should be noted that the two known LMC WN2s also have high temperatures and faint visual magnitudes (Hainich et al.\ 2014).

Their absolute visual magnitudes and spectroscopic observations rule out a massive star companion. Given our good signal-to-noise, we estimate that any companion would have to be at least 10$\times$ fainter for it to not be visible in our spectrum; such a star would have to be fainter than $M_V\sim 0$, roughly that of a B6~V. (In other words, less than 6$M_{\odot}$). Our NIR spectra further preclude there being a red giant companion. We have five to six spectroscopic observations of three of these stars and so far we have not detected any radial velocity variations, but our data are still sparse, and we are continuing to gather more.

These stars are both similar to and different from other known WNs. For example, the WN3/O3s are not the only non-binary WRs with absorption components. In the Smith et al.\ (1996) classification scheme, the spectral subtype ``WNha" is used to denote a WR star with hydrogen absorption lines (see also Crowther 2007). This nomenclature has been assigned to the Galactic stars WR3 (HD9974), WR7, WR10, WR18 and WR128, the LMC stars BAT99-18 and BATAT-63, and seven SMC WRs, all of which show WR-like emission lines with some hydrogen absorption lines (see, e.g., Martins et al.\ 2009, 2013). {\sc cmfgen} and PoWR modeling of these stars (by Marchenko et al.\ 2004, Martins et al.\ 2009, 2013, and Hainich et al.\ 2015) has constrained their physical parameters\footnote{Note: There are many other Galactic and LMC WRs with absorption lines, such as [M2002] LMC15666 (Howard \& Walborn 2012), but these have not been modeled as extensively, and have not been proven to lack O star companions.}. While there are many similarities between the physical properties of these WRha stars and the WN3/O3s, there are also a few key differences. The physical properties of the WNha stars cover a wide span of temperatures (20,000 K $< T_{\rm{eff}} < 112,000$ K), luminosities ($5.3 < \log \frac{L}{L_{\odot}} < 6.1$), helium to hydrogen ratios (0.25 to 1.2), nitrogen abundances (0.002 to 0.007), and mass-loss rates ($-5.7 < \log \dot{M} <-5.5$). Martins et al. (2009, 2013) have argued that a subset of these stars (including hydrogen-rich WNs without absorption lines, known as WNEha stars) cannot be explained by classical evolutionary models and instead appear to have evolved quasi-homogeneously. We discuss below why we believe this is not the evolutionary route for the WN3/O3s.

In terms of being different from known early-type WNs, these stars have surface He/H ratios of around 1 suggesting that they are not as chemically evolved as other early-type WNs. However, their key difference is their mass-loss rates that are 3-5$\times$ lower than what we would expect for a LMC early-type WN.

The masses we derive from the stars' surface gravities and radii are significantly smaller than what would be expected from the mass-luminosity relationship in Gr\"{a}fener et al.\ (2011), but the latter has never been checked empirically; very scant data is available from LMC WR binaries (see, e.g., Breysacher 1999). On the one hand this might suggest that these stars are more evolved (with additional mass loss during the WR stage) but this is in conflict with the He/H ratios being about 1.

In terms of binarity, Foellmi et al.\ (2003) conducted a careful radial velocity and photometric study of the LMC WN-type stars, finding only 18\% showed evidence of binarity. Correcting this for various observation effects led them to suggest that the true total binary fraction was 40\%. We have only a few radial velocity measurements of a subset of our WN3/O3s, and so the lack of binarity is not yet surprising. We also note that in most cases, the normal WNs found to have radial velocity variations have O-type companions. Because our WN3/O3s have such high temperatures, they are visually very faint ($M_V$$\sim$$-2.5$ to $-3.0$), and thus their presence might go undetected, as their optical spectra would be drowned out by their much brighter companions.

\subsection{Possible Origins}
Next we consider three possible explanations for their origin: homogeneous evolution, binarity, and single-star evolution.

As discussed above, other WNs with absorption components have been suggested to have evolved quasi-homogeneously. Such stars must have high enough initial rotational velocities ($\sim 250$ km s$^{-1}$) that material in the core mixes with material in the outer layers (Song et al. 2016). These quasi-homogeneous stars have CNO abundances typical of equilibrium, and small to medium helium enrichments (He/H = 0.25 to 0.8). Additionally Martins et al.\ (2013) found that for LMC WNs, the hydrogen-free stars had lower mass-loss rates and luminosities $\log \frac{L}{L_{\odot}} \sim 5.5$. At first glance, the WN3/O3s appear to be extreme (more evolved?) examples of these quasi-homogeneous stars. Their helium to hydrogen ratios are higher ($\sim 1$), mass-loss rates lower (-5.95), and the temperatures are on the high side ($\sim$ 100,000 K). While their enriched helium to hydrogen ratios are consistent with their high temperatures, their low mass-loss rates are more difficult to explain. F.\ Martins suggests (private communication, 2015) that the WN3/O3s might be initially more massive examples of quasi-homogeneous evolution, with lower luminosity-to-mass ratios than normal early WNs, and thus lower mass-loss rates.

There are a few arguments against these stars having evolved quasi-homogeneously. First, they have relatively low projected rotational velocities ($V_{\rm{rot}} \sim 150$ km s$^{-1}$). However, chemically homogeneous stars are generally fast rotators (time averaged surface velocity of $\sim$ 250 km s$^{-1}$; Song et al. 2016). While they may have been slowed down recently due to strong mass-loss, the low mass-loss rates of these stars make this scenario unlikely.  Indeed, comparing to theoretical predictions (which, as is expected, are still being constantly improved upon), we found that none of the homogeneously evolving LMC models of K\"{o}hler et al. (2015) can simultaneously explain both the slow rotation and the observed low mass-loss rate (D. Sz\'{e}csi, private communication, 2015). While this could be explained by the stars existing in a cluster with a metallicity that differed from the LMC, this is not the case.  As discussed earlier, we know that that the nitrogen abundances of these stars are consistent with an initial CNO abundance that is normal for the LMC. However, the stellar wind is driven primarily by Fe and we have no direct measurement; what if these stars somehow just happened to start out with low Fe abundances? Still, given that the surface nitrogen abundance is consistent with the CNO equilibrium value, the presence of hydrogen suggests a high degree of mixing between the surface and the core. However, the same composition could also be achieved by high mass loss and/or mass transfer stripping the star to reveal the core.

For all of the reasons discussed above, we are sure these stars do not have luminous massive companions. What if a punitive companion was not luminous, but rather a neutron star or black hole? We will note that, according to Vizier, none of these stars is an x-ray source, further arguing that they are not currently close binaries.  Of course, this does not preclude the possibility that a compact companion remained bound but is now in a wide orbit, and hence not generating X-rays. It might even be that the WN3/O3 was originally the more massive system, but was stripped by its putative companion which then exploded, but such a situation would occur only over a very narrow range of initial conditions, and the lifetime of the remaining star would be quite short (Pols 1994). (We are grateful to the anonymous referee for calling this scenario to our attention.)  We also consider the possibility that at some previous point the WN3/O3s were part of a binary system that has since merged. For instance, one could merge a WNE star with a lower mass main-sequence star and perhaps get something like what we see. Still, one would imagine that as the cores merge and the envelopes mix, some material would be ejected, and the resulting star would be a rapid rotator -- which none of our stars are. And, these scenarios become more improbable given that we have 9 star total, or 8\% of the LMC's WNs. Furthermore, all of these stars have radial velocities consistent with their location within the LMC according to the kinematic models created by Olsen \& Massey (2007). This suggests that if they originally had a more massive companion that exploded, leaving behind a neutron star or black hole remnant, the remnant should either still be there (likely generating X-rays) or have merged (causing high rotation).

That leaves us with the possibility that these stars are in fact just a normal short-lived phase in normal single star evolution. Given the He/H ratio, these stars might be intermediate between O stars and WNs, and that they will progress to ``normal" WNs with denser winds as they lose more of their hydrogen envelopes. Their bolometric luminosities are typical for other WNs in the LMC, as shown in Figure~\ref{fig:tefflum}, arguing, perhaps, that these are not lower mass versions of WNs, but more normal.  The fact that these stars are relatively common in the LMC, but not seen in the Milky Way, could then be a consequence of the the fact that the mass-loss rates have a strong dependence on the Eddington factor (Gr\"{a}fener et al. 2011). The lower metallicity of the LMC might affect the strengths and duration of the stellar winds.

\begin{figure}
\plotone{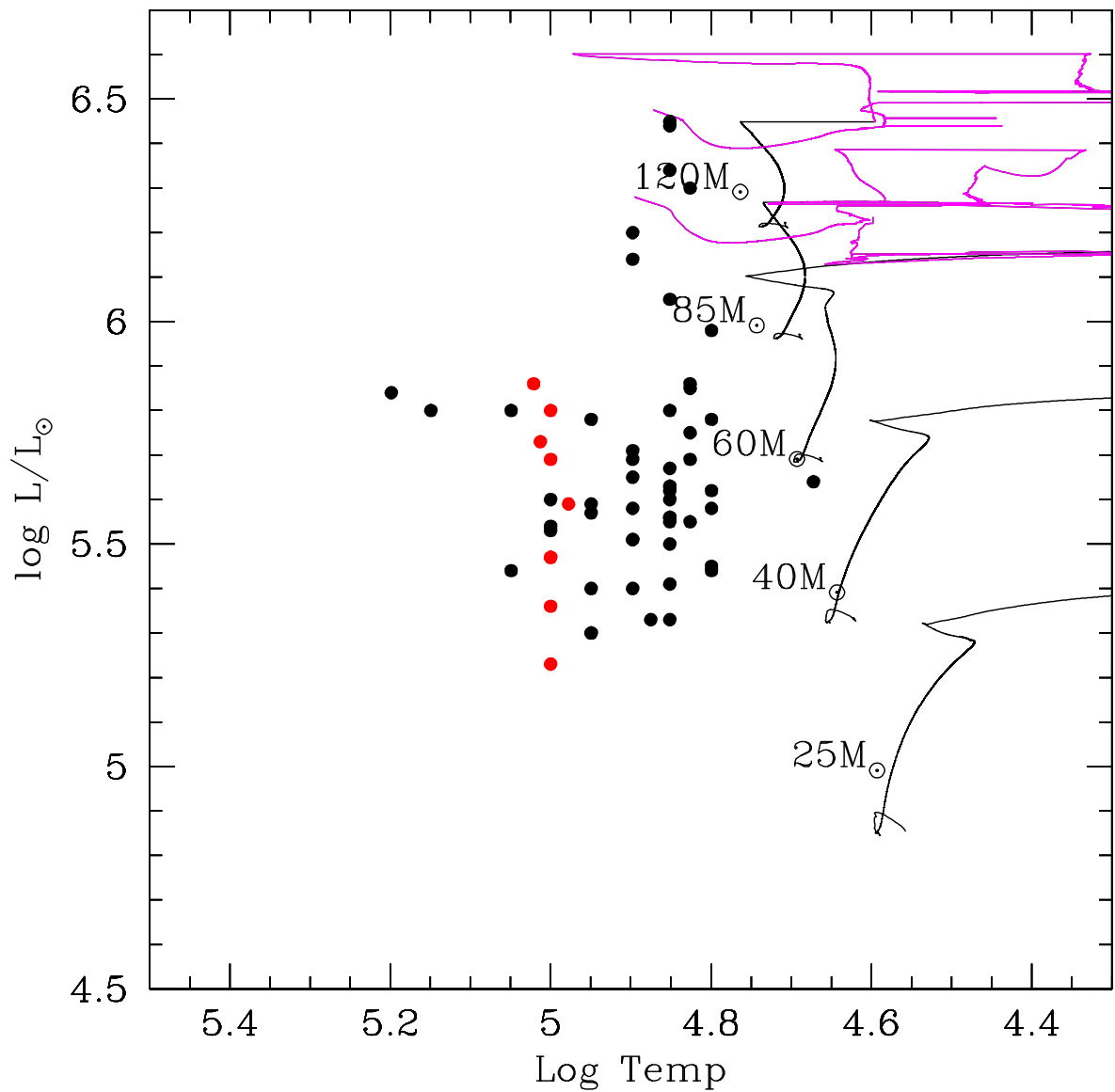}
\caption{\label{fig:HRD} The location in the HRD of the 9 WN3/O3s (red dots) from this study and the early WNs (black dots) from Hainich et al. (2014).  The preliminary version of the Geneva evolution tracks for Z=0.006 are shown;  these models include rotation and are an updated version to those used by Neugent et al.\ (2012b); for a description; see Ekstr\"{o}m et al.\ (2012). The magenta part of the tracks indicate the WR phase (i.e., hydrogen mass fraction $<$0.3 and temperature above 10,000 K, following Georgy et al.\ 2012).}
\end{figure}

In Figure~\ref{fig:HRD} we show the location of our WN3/O3s in the HRD along with the latest Geneva evolutionary models with a metallicity corresponding to that of the LMC (Z=0.006). These models include the effect of rotation (initial velocities of 40\% of the break up speed), and are an updated version of the models used by Neugent et al.\ (2012b) in their study of yellow and red supergiants in the LMC. These models are still undergoing refinement prior to final publication, and we are grateful to C.\ Georgy for sharing these with us. We see that these models fail to reproduce WR stars with luminosities as low as those of the early-type WNs, either our WN3/O3s or those in the Hainich et al. (2014) sample. The lowest luminosity WRs predicted by the tracks have $\log L/L_\odot \sim$6. The evolved parts of the tracks also do not extend to high enough effective temperatures, but that is less worrisome, as it depends critically on the physical treatment of the outer layers of the star, but the luminosity is fundamentally tied to the total mass of the star and its internal physics; see discussion in Georgy et al. 2012. The older Geneva tracks (Meynet \& Maeder 2005) used for comparison by Hainich et al.\ (2014) in their Figure 10 did extend to lower luminosities ($\log L/L_\odot \sim$5.5) and higher temperatures, but there have been numerous improvements in the models in the intervening decade; see Ekstr\"{o}m et al.\  (2012). However, a similar problem (although not as severe) is encountered at Milky Way metallicities; see Figure 3 in Georgy et al. (2012). One possible solution to this quandary would be that the mass-loss rates during the RSG phase are significantly underestimated in the models (see, e.g., Maeder et al. 2015). Possibly this would allow stars of lower mass and luminosity to evolve to WRs.

\section{Conclusions and Next Steps}
The properties of these stars have raised an interesting possibility as to their ultimate fate.  Drout et al.\ (2016) recently studied various types of supernovae progenitors and found that WN3/O3s, with their low mass-loss rates and high terminal wind velocities, could be the progenitors to the Type Ic-BL supernovae (see Drout et al.'s Figure 17). A subset of these Type Ic-BL supernovae then turn into long-duration gamma ray bursts which are preferentially found in low metallicity environments (Vink et al. 2011). Thus, considering that WN3/O3s might be the progenitors to such long duration gamma ray bursts, we are curious whether there is a metallicity dependence for their formation.

\begin{figure}
\epsscale{1.0}
\includegraphics[width=3.2in]{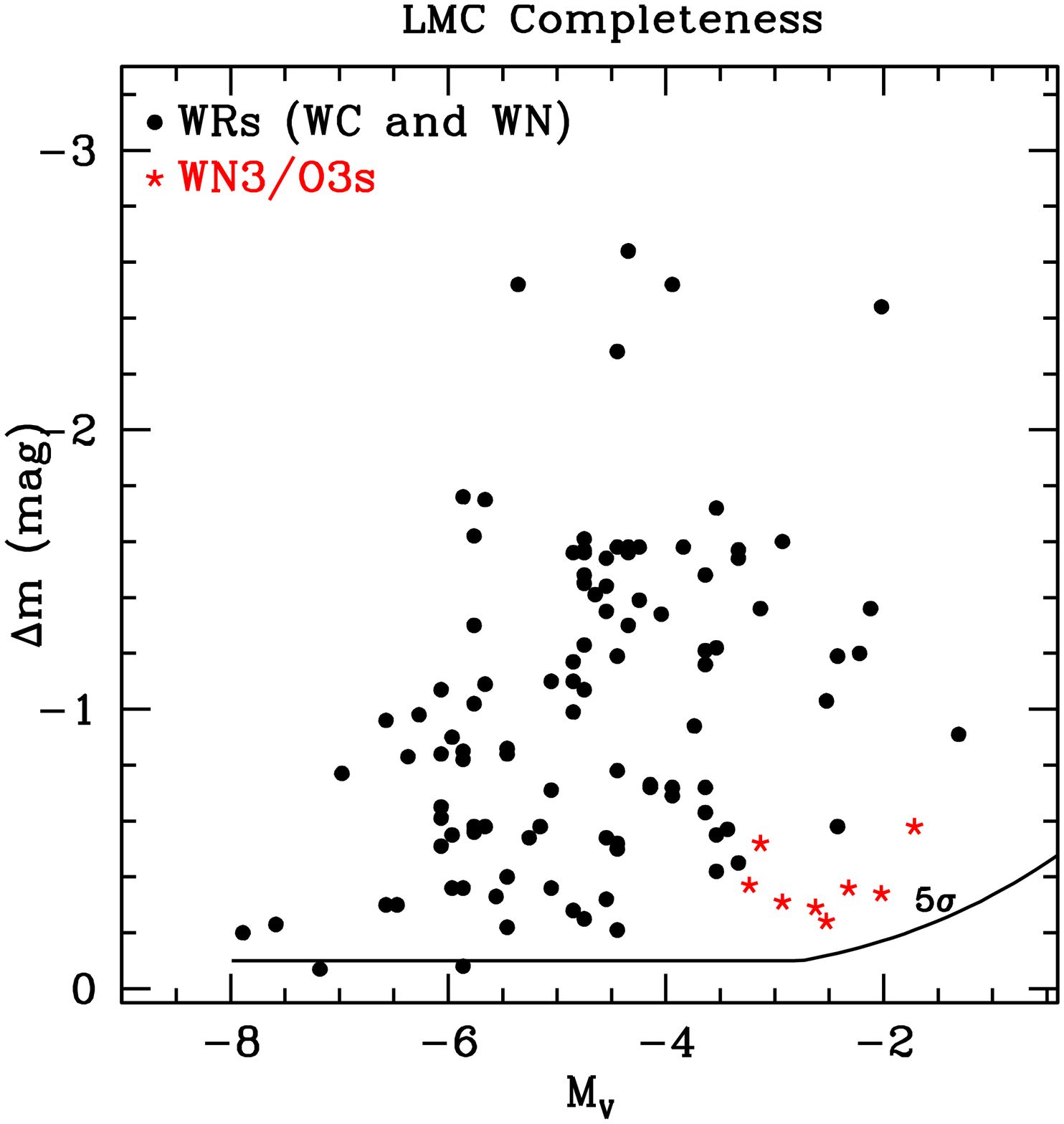} 
\includegraphics[width=3.2in]{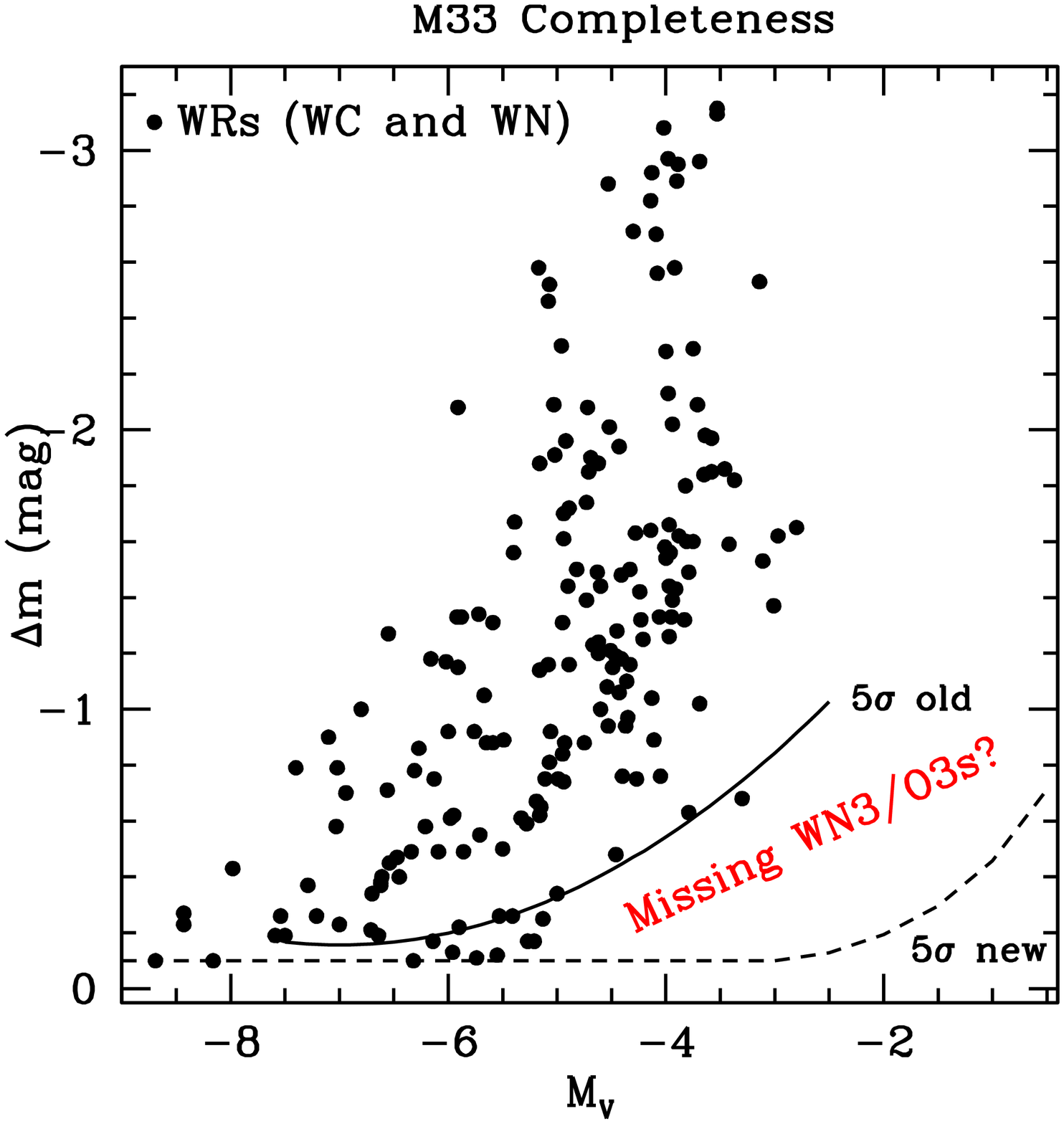} 
\caption{\label{fig:M33} Magnitude difference (emission minus continuum filter) vs.\ absolute magnitude showing that our previous M33 surveys did not go deep enough to detect WN3/O3s.}
\end{figure}

So far, the WN3/O3s have only been found in the LMC even though hundreds of WRs have been found in the Milky Way. While there are other WRs in the Milky Way with absorption lines, none of them are very similar to the WN3/O3s. Given this, we expect there might be a metallicity dependence on their formation. To test this, we have turned to M33, which has a strong metallicity gradient with log(O/H) + 12 = 8.7 at the center and log(O/H) + 12 = 8.3 in the outer regions (see Magrini et al. 2007). While Neugent \& Massey (2011) searched for WRs in M33, their survey did not go deep enough to detect these faint stars, as is shown in Figure~\ref{fig:M33}. Thus there may be an entirely unexplored population of WN3/O3s in M33's lower metallicity regions. Hopefully our future survey will either prove or deny the existence of these WN3/O3s in different metallicity environments and help us understand more about this new type of WR.

\acknowledgements
We acknowledge valuable conversations and correspondence with many of our colleagues, including Paul Crowther, Cyril Georgy, Jose Groh, and Georges Meynet. We thank Knut Olsen for calculating the expected radial velocities of our nine stars based on kinematic modeling of the LMC and Cyril Georgy for the preliminary LMC-like metallicity evolutionary tracks. Finally, we appreciate the helpful comments of the anonymous referee that strengthened and improved the paper. This work has been partially supported by the National Science Foundation under AST-1008020. Support for program number HST-GO-13780 was provided by NASA through a grant from the Space Telescope Science Institute, which is operated by the Association of Universities for Research in Astronomy, Incorporated, under NASA contract NAS5-2655. D.J.H. also acknowledges support from STScI theory grant HST-AR-12640.01. We would additionally like to thank the excellent support staff at Las Campanas for all of their help during our observing runs.

\clearpage

\clearpage
\begin{deluxetable}{l c c c c c c c }
\tabletypesize{\scriptsize}
\tablecaption{\label{tab:observations} Summary of Observations}
\tablewidth{0pt}
\tablehead{
Region:
& \multicolumn{2}{c}{UV}
&
& \multicolumn{2}{c}{Optical}
&
& \multicolumn{1}{c}{NIR} \\  \cline{2-3} \cline{5-6} \cline{8-8}
Instrument:
& \multicolumn{2}{c}{HST/COS}
&
&\colhead{MagE}
&\colhead{MIKE}
&
& \colhead{FIRE} \\   \cline{2-3}
Disperser:
&\colhead{G140L/1105}
&\colhead{G140L/1280}
&
&\colhead{Fixed}
&\colhead{Fixed}
&
&\colhead{Echelle} \\
Wavelength:
&\colhead{1112-2000 \AA}
&\colhead{920-2000\tablenotemark{a}\AA}
&
&\colhead{3180-9400 \AA}
&\colhead{3300-8900 \AA}
&
&\colhead{0.83-2.08 $\mu$m} \\
Resolving Power:
&\colhead{1500-4000}
&\colhead{1500-4000}
&
&\colhead{4100}
&\colhead{12000}
&
&\colhead{6000} \\ \cline{1-8}
Star
&\multicolumn{7}{c}{Exposure times (sec)\tablenotemark{b}}
}
\startdata
LMC079-1 & 2405 & 2899 & &3600(4) & \nodata && \nodata \\
LMC170-2 & 2388 & 2856 & &3900(4) & 1800 && 2400 \\
LMC172-1 & \nodata & \nodata && 3600(3) & \nodata &&\nodata \\
LMC174-1 & \nodata & \nodata && 1000(1) & \nodata && \nodata \\
LMC199-1 & \nodata & \nodata &&2400(2) & \nodata && \nodata\\
LMC277-2 & 2388 & 2857 && 1800(2) & \nodata && \nodata \\
LMCe078-3 & \nodata & \nodata && 1200(1) & \nodata && \nodata \\
LMCe159-1 & \nodata & \nodata && 1200(1) & \nodata && \nodata \\
LMCe169-1 & \nodata & \nodata && 1800(1) & \nodata && \nodata \\
\enddata
\tablenotetext{a}{With a gap from 1185-1270\AA.}
\tablenotetext{b}{The number in parenthesis indicate the number of MagE spectra.}
\end{deluxetable}

\begin{deluxetable}{l l r r}
\tablecaption{\label{tab:photometry} Observational Properties of the WN3/O3s}
\tablewidth{0pt}
\tablehead{
\colhead{Star}
& \colhead{V\tablenotemark{a}}
& \colhead{B-V\tablenotemark{a}}
& \colhead{M$_{\rm{V}}$\tablenotemark{b}}
}
\startdata
LMC079-1 & 16.31 & -0.25 & -2.8\\
LMC170-2 & 16.13 & -0.17 & -2.9\\
LMC172-1 & 15.95 & -0.12 & -3.0\\
LMC174-1 & 17.11 & 0.14 & -3.0\\
LMC199-1 & 16.65 & -0.22 & -2.3 \\
LMC277-2 & 15.83 & -0.16 & -3.1 \\
LMCe078-3 & 17.12 & -0.19 & -2.2 \\
LMCe159-1 & 16.15 & -0.10 & -2.6\\
LMCe169-1 & 17.03 & 0.03 & -1.8\\
\enddata
\tablenotetext{a}{Values from Zaritsky et al., 2004}
\tablenotetext{b}{Values from Massey et al., 2014; 2015; 2017}
\end{deluxetable}

\begin{deluxetable}{l l l l}
\tablecaption{\label{tab:params} {\sc cmfgen} model parameters for LMC170-2 based on optical spectra}
\tablewidth{0pt}
\tablehead{
\colhead{Parameter}
& \colhead{Best Fit}
& \colhead{Uncertainty\tablenotemark{a}}
}
\startdata   
$T_{\rm{eff}}$ (kK) & 100 & 5\\
$\log \frac{L}{L_{\odot}}$ & 5.60 & 0.1\\
$\log \dot{M}$\tablenotemark{b} & -5.95 & 0.05 \\
$\log g_{\rm eff}$ & 4.95 & 0.05\\
He/H (by \#) & 1.0 & 0.2\\
N (by mass) & 0.011 & -\\
C (by mass) & $1.5 \times 10^{-4}$ & -\\
O (by mass) & $4.8 \times 10^{-4}$ & -\\
$v_\infty$ (km s$^{-1}$) & 2400 & 200\\
$v_{\rm sin i}$ (km s$^{-1}$) & 150 & 10\\
$\beta$ & 0.8 & -
\enddata
\tablenotetext{a}{A dash means we were unable to estimate the uncertainty}
\tablenotetext{b}{Assumes a clumping filling factor of 0.1}
\end{deluxetable}

\begin{deluxetable}{l l l l l l l l l}
\tabletypesize{\scriptsize}
\tablecaption{\label{tab:allParams} {\sc cmfgen} model parameters for nine new WN3/O3s \tablenotemark{*}}
\tablewidth{0pt}
\tablehead{
\colhead{Star}
& \colhead{$T_{\rm{eff}}$ (kK)}
& \colhead{$\log \frac{L}{L_{\odot}}$}
& \colhead{$\log g$ [cgs]}
& \colhead{$\log \dot{M}$}
& \colhead{He/H (by \#)}
& \colhead{N (by mass)}
& \colhead{Radius (R$_{\odot}$)}
& \colhead{Mass (M$_{\odot}$)}}
\startdata
LMC079-1 & 95 & 5.59 & 4.85 & -6.00 & 0.7 & 0.0055 & 2.1 & 11\\
LMC170-2 & 100 & 5.69 & 4.95 & -5.82 & 1 & 0.011 & 2.2 & 15\\
LMC172-1 & 105 & 5.86 & 4.95 & -5.66 & 1.5 & 0.0055 & 2.3 & 16\\
LMC174-1 & 100 & 5.47 & 4.95 & -5.66 & 1 & 0.0055 & 1.7 & 9\\
LMC199-1 & 100 & 5.47 & 4.95 & -5.82 & 1 & 0.011 & 1.7 & 9\\
LMC277-2 & 100 & 5.80 & 4.95 & -5.85 & 1 & 0.0055 & 2.5 & 19\\
LMCe078-3 & 100 & 5.36 & 4.95 & -6.04 & 1 & 0.011 & 1.5 & 7\\
LMCe159-1 & 103 & 5.73 & 4.95 & -5.98 & 1 & 0.0055 & 2.1 & 14\\
LMCe169-1 & 100 & 5.23 & 4.95 & -5.85 & 1 & 0.011 & 1.3 & 6\\
\enddata
\tablenotetext{*}{For all nine models: C = 1$\times10^{-4}$ (by mass), O = 8$\times10^{-5}$ (by mass), $v_{\infty} = 2600$ km s$^{-1}$, $v_{\sin i} = 150$ km s$^{-1}$, $\beta = 1$, and the clumping filling factor = 0.1. The uncertainties are the same as stated in Table 3.}
\end{deluxetable}

\end{document}